\documentclass[prx,twocolumn,superscriptaddress]{revtex4}
\usepackage{natbib}
\usepackage{amsmath}
\usepackage{amsfonts}
\usepackage{graphicx}
\usepackage{dcolumn}
\usepackage{tabularx}
\usepackage{hyperref}
\usepackage{keyval}
\usepackage{multirow}
\usepackage{xcolor}
\usepackage{CJKutf8}

\begin{document}
\begin{CJK*}{UTF8}{bsmi}

\title{Strange metal dynamics across the phase diagram of Bi$_{2}$Sr$_{2}$CuO$_{6+\delta}$ cuprates.}
\author{Erik van Heumen}\email{e.vanheumen@uva.nl}
\affiliation{van der Waals - Zeeman Institute, University of Amsterdam, Sciencepark 904, 1098 XH Amsterdam, the Netherlands}
\affiliation{QuSoft, Science Park 123, 1098 XG Amsterdam, The Netherlands}
\author{Xuanbo Feng (馮翾博)}
\affiliation{van der Waals - Zeeman Institute, University of Amsterdam, Sciencepark 904, 1098 XH Amsterdam, the Netherlands}\affiliation{QuSoft, Science Park 123, 1098 XG Amsterdam, The Netherlands}
\author{Silvia Cassanelli}
\author{Linda Neubrand}
\author{Lennart de Jager}
\author{Maarten Berben}
\author{Yingkai Huang}
\affiliation{van der Waals - Zeeman Institute, University of Amsterdam, Sciencepark 904, 1098 XH Amsterdam, the Netherlands}
\author{Takeshi Kondo}
\affiliation{Institute for Solid State Physics, University of Tokyo, Kashiwa-no-ha, Kashiwa, Japan}
\author{Tsunehiro Takeuchi}
\affiliation{Toyota Technological Institute, Nagoya 468-8511, Japan}
\author{Jan Zaanen}\email{jan@lorentz.leidenuiv.nl}
\affiliation{Lorentz institute, Universiteit Leiden, Leiden, the Netherlands}

\begin{abstract} 
Unlocking the mystery of the strange metal state has become the focal point of high T$_{c}$ research, not because of its importance for superconductivity, but because it appears to represent a truly novel phase of matter dubbed `quantum supreme matter'. Detected originally through high magnetic field, transport experiments, signatures of this phase have now been uncovered with a variety of probes. Our high resolution optical data of the low T$_{c}$ cuprate superconductor, Bi$_{2-x}$Pb$_{x}$Sr$_{2-y}$La$_{y}$CuO$_{6+\delta}$ allows us to probe this phase over a large energy and temperature window. We demonstrate that the optical signatures of the strange metal phase persist throughout the phase diagram. The strange metal signatures in the optical conductivity are two-fold, (i): a low energy Drude response with Drude width on the order of temperature and (ii): a high energy conformal tail with doping dependent power-law exponent. While the Drude weight evolves monotonously throughout the entire doping range studied, the spectral weight contained in the high energy conformal tail appears to be doping and temperature independent. Our analysis further shows that the temperature dependence of the optical conductivity is completely determined by the Drude parameters. Our results indicate that there is no critical doping level inside the superconducting dome where the carrier density starts to change drastically and that the previously observed 'return to normalcy' is a consequence of the increasing importance of the Drude component relative to the conformal tail with doping. Importantly, both the doping and temperature dependence of the resistivity are largely determined by the Drude width.
\end{abstract}

\maketitle
\end{CJK*}

\section{Introduction}
Since the discovery of cuprate high Tc superconductivity thirty-five years ago, numerous studies of the optical properties were published. It may appear as an exhausted affair -- everything that could be measured has been measured -- and this reflected in the reduced output over the last ten years or so \cite{Bilbro:natphys2011, Homes:PRB2012, Mirzaei:PNAS2013, Moon:PRB2014, Levallois:PRX2016, Ohnishi:JPSJ2018, mahmood:PRL2019,Lyzwa:PRR2020, Michon:PRR2021}. Nevertheless, high T$_{c}$ superconductivity continues to be a fertile source of surprises \cite{Keimer:nature2015}. If anything, the profundity of the mystery pointing at a fundamentally different type of physics has become more manifest in recent years. One aspect is that the theoretical view on the physics behind transport phenomena has been on the move. During the early history of the subject it was taken for granted that the transport originates in a very dilute gas of thermally excited quasiparticles as in conventional Fermi-liquid metals. Catalyzed by the understanding of strongly interacting quantum critical states of matter \cite{Sachdev:book2011} and further elaborated by insights coming from the AdS/CFT correspondence of string theory \cite{Zaanen:book2015, Hartnoll:book} it was realized that in his regard Fermi-liquids are singularly special. Yet other states of strongly interacting quantum matter may be formed, characterized by dense many body quantum entanglement \cite{Zaanen:arxiv2021}. Rooted in the advancements in the general understanding of quantum (Eigenstate) thermalization \cite{Liu:RPP2020} one expects that such non-Fermi-liquids are characterized by extremely rapid thermalization and the absence of quasiparticle excitations \cite{Zaanen:arxiv2021}. One then expects that the transport in such systems eventually rests on the highly collective flows described by hydrodynamics, or otherwise in the form of `incoherent' transport. The latter should reflect simple scaling properties related to what is found at thermal phase transitions. 

On the experimental side, doubts regarding the Fermi liquid arose early in the form of the famous linear-in-temperature DC electrical resistivity, $\rho (T) \sim T$. This linearity of the resistivity in optimally doped cuprates extends all the way from the superconducting $T_c$ up to the melting point of the crystal, with $\rho$ becoming larger than what is expected from the Mott-Ioffe-Regel minimal conductivity criterium expected in normal metals. Given the high T$_{c}$ of these materials, one could argue that the normal state is always in the phonon scattering dominated regime. However, this linearity extends to sub-kelvin temperatures when superconductivity is suppressed in high magnetic fields\cite{Cooper:science2009}. The problem of principle has been all along to explain why this behaviour is so simple -- dealing with quasiparticles the resistivity should be a more interesting function of temperature. Recently the `Planckian dissipation' \cite{Zaanen:nat2004} came into the limelight \cite{Hartnoll:natphys2015, Lucas:PRB2018}. Taking for granted that the DC transport is of the Drude kind it was deduced that the current relaxation (scattering) time $\tau_J \simeq \tau_{\hbar}$. The Planckian dissipation time $\tau_{\hbar} = \hbar / (k_B T)$ is the quantum physical time scale associated with dissipative physics at finite temperature which appears naturally in the context of the quantum thermalization of densely entangled states of matter \cite{Zaanen:SciPost2019}. Claims appeared recently that it is a remarkably ubiquitous property, showing up also in a variety of non-cuprate strange metal systems \cite{Hartnoll:arXiv2021}. 

In this recent era the focus in cuprate research shifted to the overdoped regime where especially DC magneto-transport experiments revealed yet other anomalous properties. It was claimed that the carrier density jumps discontinuously near optimal doping from a semi-conductor behaviour $\sim x$ (doping) to the Luttinger volume $\sim 1 + x$ associated with a large Fermi surface \cite{Badoux:Nature2016}. Other groups subsequently reported that there may not be a jump \cite{Putzke:nphys2021}. Regardless, the carrier density in the overdoped regime as revealed by recent Hall measurements develops in an anomalous way suggestive of two charge reservoirs existing in parallel, a notion getting further support by a highly anomalous magneto-resistance \cite{Ayres:nature2021}. Contrary to the long standing belief that Fermi-liquid physics resurrects beyond optimal doping, the evidence is now mounting that the overdoped metallic state is yet another theatre to study the `strange' physics. 

Compared to DC transport, AC transport reveals much more information regarding the fundamentals of transport. By linear response principles, this reveals the optical conductivity $\hat{\sigma} (\omega, T) = \sigma_1 (\omega, T) + i \sigma_2 (\omega, T)$. The dissipative ($\sigma_1(\omega)$) and reactive ($ \sigma_2(\omega)$) parts, related to each other through causality, can both be measured and the DC conductivity corresponds with no more than the zero frequency asymptote $\sigma_1 (\omega =0, T)$. The essence of what follows is that the {\em analytical properties} of the complex function $\hat{\sigma}(\omega, T) $ are a rich source of quintessential {\em phenomenological} information. Having even not the faintest clue regarding the microscopic physics one can extract some tight bounds that are eventually rooted in symmetry principles. 

Already when the first optical conductivity data appeared in the late 1980's \cite{Thomas:PRB1988,Collins:PRB1989} the large peak centred at zero frequency observed in the strange metal regime was assigned immediately to the Drude response, ubiquitous in normal metals. It became gradually clear that the width of this peak (the current relaxation rate) $\sim 1/\tau_{\hbar}$ \cite{Quijada:PRB1999, Marel:nature2003}, while the Drude weight is to good approximation temperature independent. Since $\rho (\omega = 0, T)\,=\,(\omega^2_p\tau_J)^{-1}$ in a Drude conductor, this constitutes the evidence for the Planckian momentum life time. 

In text books this Drude response is typically tied to quasiparticles -- by reference to the Sommerfeld model -- and it was conceptualized like this in this early era. However, such a Drude response is actually completely generic for {\em any} finite density charged fluid living in a spatial manifold characterized by a weak translational symmetry breaking \cite{Zaanen:SciPost2019}. It just reflects the fact that the total momentum of the fluid is long lived (see Section \ref{Gentransport}). For instance, the `unparticle' fluids of AdS/CFT also exhibit a Drude response under such circumstances \cite{Horowitz:JHEP2012,Donos:JHEP2015}.

There is however excess spectral weight at higher frequencies coined `mid-infrared absorptions' (or generalized Drude) \cite{Reedyk:PRB1988, Romero:PRL1992, Quijada:PRB1999}. However, in Ref. \cite{Marel:nature2003,Marel:AP2006} it was shown that conductivity response function in the energy range $0.1 < \omega < 1$ eV has a special analytical form: it follows the branch-cut,
\begin{equation}
\sigma (\omega) \sim 1 / (i \omega )^{\alpha} 
\label{Eq:branchcut}
\end{equation} 
 where the `anomalous scaling dimension' $\alpha\approx\,0.7$ at optimal doping in the bilayer BSSCO system. This demonstration demands the experimental availability of both $\sigma_1(\omega)$ and $\sigma_2(\omega)$; next to the fall-off according to $ | \sigma (\omega) | \sim |\omega |^{-\alpha_{\sigma}}$ the property of the branch cut that the phase angle is frequency independent and set by $ \alpha \pi /2$ has to be reflected in the data and this requires knowledge of both $\sigma_1(\omega)$ and $\sigma_2(\omega)$. Such a branch cut is a typical scaling form associated with some form of quantum criticality -- it has been nicknamed the `conformal tail' referring to conformal invariance. 

But to which degree is the view that we just sketched accurate? Are the transport phenomena in these strange metals reflected by hydrodynamical principle as suggested by the recent theoretical advances? Is it rooted in quasi-particles according to the long standing belief, or is it yet something else? Given the gravitas of this affair, we do find that the data deserve an interpretational scrutiny reminiscent of the standards in for instance high energy physics or cosmology. Instead of attempting to use the data to fit it with the expectations originating in a particular theory or model, one departs from generic constraints associated with the measurement process to arrive at {\em bounds} pertaining to the interpretational models. Which information can be extracted from the data that relates to a particular interpretation, and what are the limitations? 

Resting on strictly phenomenological means independent of theoretical notions, here we wish to make a start with such an endeavor inviting the readership to arrive at further improvements. Besides general requirements of symmetry, sum rules and causality (Kramers-Kronig consistency) we exploit the analytical properties of $\sigma (\omega)$ as sketched in the above to aim at a confidence level in the interpretation in terms of bounds we can extract from the data. The outcomes are surprising, to a degree adding precision to the conventional view but also showing that these contain flaws. 

In addition, in light of the present interest in the overdoped regime, we ask how the optical functions evolve as function of overdoping. We focus on the single layer bismuth cuprates (Bi$_{2}$Sr$_{2}$CuO$_{6+\delta}$, BSCO), because it is comparatively easy to change the carrier density in the overdoped regime through oxygen or vacuum annealing. Using high quality reflectivity experiments, and for some crystals additional ellipsometry data, we can reconstruct with high fidelity both $\sigma_1$ and $\sigma_2$ over a frequency range of $0.01$ to $\sim 4$ eV, a temperature range $10-300$ K and a doping range spanning the phase diagram from non-superconducting, underdoped to non-superconducting overdoped crystals ($p = 0.05 - 0.28$). 

Our findings are as follows. 
\begin{itemize}
\item[1.] Is the low energy (including DC) transport actually of the Drude kind? Resting on a high precision analysis we will derive in Section \ref{nofrillsDrude} an upper bound for {\em non-Drude} contributions to the DC transport being $\lesssim10 \%$ over the whole doping range, up to room temperature. We confirm that the low frequency momentum relaxation rate (the Drude width) is of Planckian magnitude. As is impossible from the DC transport measurements, we can unambiguously determine the carrier density from the data in the form of the Drude weight. We find this to increase in a smooth, linear way from the slightly underdoped- to the strongly overdoped regime. There is no sign of any irregularity at the critical doping $p_c$. Claims of the kind in this regard based on the Hall effect are therefore based on a flawed interpretation. 

\item[2.] How does the conformal tail evolve with doping? Its fingerprint is the algebraic fall off setting in at $\omega \gtrsim 200$ meV, characterized by the constancy of the phase angle as discussed in the above. We will show in Section \ref{branchcut} that this {\em persists over the full doping range} into the strongly overdoped regime. The spectral weight associated with this conformal tail is rather doping independent, demonstrating that in this regard even at the very high doping level where the superconductivity has disappeared the metals continue to exhibit unconventional behaviour, at least in this regard. Most significantly, we find that the anomalous dimension $\alpha_{\sigma}$ shows a considerable variation as function of doping, reminiscent of what is expected in a quantum critical {\em phase} of matter. Yet again, nothing special is found at the doping $p_c$ where the putative quantum phase transition should reside. 

\item[3.] That there is a high energy (conformal tail) and low energy (Drude) regime is beyond doubt, but how to characterize the crossover around $100$ meV between these distinct behaviours? Here we face an ambiguity, the question being whether these should be viewed as conductors in series or in parallel. We will explore both scenario's. In the first case one should add up the conductivities of both sectors and this turns out to be tightly constrained by the requirement that both $\sigma_1$ and $\sigma_2$ as related by Kramers-Kronig should be reproduced. As we will show in Section \ref{AntiMatcrossover} a consistent fit can be obtained by a form where the incoherent part terminates at a broadened gap, Eq. (\ref{AntiMatfit}). On the other hand, for the `in series' case a single optical self-energy (memory function) will be in effect (Section \ref{Matthiessencrossover}). This reveals equally well the presence of the crossover energy scale, resting on the fact that this self-energy contains the information regarding the {\em decoupling} of the electrical current from momentum upon entering the conformal tail regime. This however introduces subtle ambiguities for the physics. In the second interpretation, the momentum relaxation rate acquires a frequency dependence in the Drude regime that may be of the type envisaged in e.g. the marginal Fermi-liquid phenomenology. However, in the first scenario an energy dependence of $1/\tau_P$ cannot be resolved since the deviations of a simple Drude could well be entirely due to the spilling over of the incoherent spectral weight due to the gap smearing.

\end{itemize}

This summarizes our main results. The organization of the paper is as follows, we will first present a short reminder of the gross principles underlying transport in Section \ref{Gentransport}, followed by a short overview of the experimental optical response (Section \ref{exp_details}, Appendix \ref{appA}). We will then continue analyzing the various spectral regimes: the low energy Drude response (Section \ref{nofrillsDrude}), the high energy branch cut part (Section \ref{branchcut}), and the crossover regime (Section \ref{crossoverreg}). In Section \ref{discconc} we will discuss the ramifications of these results for the various theoretical proposals. 

\section{Generalities of transport: a reminder.}
\label{Gentransport}

Before we turn to the data analysis it may be beneficial for some readers to get reminded of the highly generic nature of a Drude response. Given the way it is taught in elementary solid state courses one may have gotten the impression that this type of response is special for an extremely dilute gas formed from quasi-particles loosing individually momentum by scattering from obstacles -- the Drude-Sommerfeld affair. This is mere folklore -- although accurate for the transport in normal metals it is in fact in the general context of the dynamics of fluids a quite pathological limit. 

What is Drude transport about? One is actually measuring the response of the {\em macroscopic} charged fluid sourced by an electrical field. This sets an electrical current in motion and all one has to assert is that this current $\vec{J}$ lives for a finite time $\tau_J$. Assuming a charge density $n e$ and a microscopic (band) mass $m_b$ this results in the simple equation of motion,

\begin{equation}
\frac{\partial \vec{J}}{\partial t} + \frac{1}{\tau_J} \vec{J} = \frac{ne^{2}}{m_{b}} {\vec E} 
\label{EOMmomentum}
\end{equation}

Defining the longitudinal conductivity $\sigma$ through $\vec{J} = \sigma \vec{E}$ and invoking a source oscillating at frequency $\omega$ one finds for the conductivity in full generality,

\begin{equation} 
\sigma (\omega) = \frac{\omega^2_p}{4 \pi} \frac{i}{ ( \omega + M (\omega) )}
\label{Memfunction}
\end{equation}

This is the Drude optical response, characterized generically by an overall Drude weight parametrized here in terms of the plasma frequency $\omega_p = \sqrt{ n e^2 / m_b}$ where $n$ is the carrier density. In full generality, $M (\omega) = M_1 (\omega) + i M_2 (\omega)$ is the memory function (also called the optical self energy), a Kramers-Kronig consistent complex function. Next to energy- and charge conservation, the breaking of translational invariance plays a key role. This current may partially overlap with  the {\em total} macroscopic momentum $\vec{P}_{\mathrm{tot}}$: $\vec{J} = n e \vec{P}_{\mathrm{tot}}/ m_b$. In the Galilean continuum for finite rest mass, this overlap is complete as $\vec{P}_{\mathrm{tot}}$ is the conserved Noether charge associated with the continuous space translations. The optical conductivity in the continuum will be a delta function at zero energy describing the perfect metal. Upon breaking translational symmetry in a solid, the spectral weight contained in the delta function will redistribute to finite frequency, involving for instance interband transitions or the intraband response of Fermi-liquids associated with the dissipation of the current, $\tau_J$. Dealing with a simple relaxational response one identifies $M_2 = 1/\tau_J$ (and $M_1 = 0$) to find,
\begin{eqnarray} 
\sigma (\omega) & = & \frac{\omega^2_p \tau_P}{4 \pi} \frac{1}{ 1 - i (\omega \tau_J)} \nonumber \\
\sigma_1 (\omega) & = & \frac{\omega^2_p \tau_P}{4 \pi} \frac{1}{ 1 + (\omega \tau_J)^2}
\label{Simpledrude}
\end{eqnarray}

coincident with the response that follows directly from the simple EOM, Eq. (\ref{EOMmomentum}). This is the Drude conductivity quoted in the texbooks. It follows immediately that the DC conductivity $\sigma (\omega = 0) = \omega_p^2 \tau_J/4\pi$: the combination of Drude weight and the momentum relaxation time is measured in transport experiments. To obtain these quantities separately one has to inspect the finite frequency response. This is the `half-Lorentzian' Eq. (\ref{Simpledrude}) with a Drude-weight set by the total area, while the Drude-width reveals $1/\tau_J$. 

As spelled out by the Mori-Zwanzig formalism this generalizes to a frequency dependent memory function when one is dealing with a multitude of relaxation times. For instance, in a Fermi-liquid exposed to an Umklapp potential the momentum relaxation is set by the quasiparticle collision rate resulting in $M_2 (\omega, T) \sim (1 / \hbar E_F) ( (\hbar \omega)^2 + (2\pi k_B T)^2)$ \cite{Berthod:PRB2013}. 

The take home message is that a Drude response is generic. One measures the motion of the {\em macroscopic} ($ q \rightarrow 0$) fluid at a finite temperature. When $\hbar \omega << k_B T$ (i.e., the DC measurement) the thermal fluid formed at finite temperature has to be governed by classical stochastic dynamics. At a finite density the electrical currents will overlap with momentum for any fluid and momentum life time is then the universal limiting factor. The information regarding the microscopic physics enters via the parameters of the macroscopic theory -- the Drude weight and width -- but also in the frequency dependence of $\sigma (\omega)$ when $\hbar \omega > k_B T$. In conventional metals the momentum relaxation times are directly linked to the {\em single particle} momentum life times but this is actually a highly special affair linked to the extreme dilute gas circumstances rooted in the zero temperature Fermi liquid microscopics. Given the now rather well understood quantum thermalization mechanisms operative in the densely entangled non-Fermi liquids, the expected rapid thermalization will render the macroscopic fluid to be ruled by Navier-Stokes hydrodynamics even in the presence of substantial disorder. This is then characterized by unusual hydrodynamical parameters such as the `minimal viscosity' \cite{Zaanen:SciPost2019}. The punchline is that upon interrogating such a fluid with optical means one will see the same typical Drude peak as expressed in Eq. (\ref{Simpledrude}). The only meaningful question that remains is how the Drude parameters depend on the physical circumstances.

As a final caveat, dealing with very special circumstances such as charge conjugation symmetry (e.g., graphene at zero density) or a diverging momentum susceptibility \cite{Else:PRL2021} the electrical current can decouple from total momentum and this will show as an incoherent response. Dealing with whatever form of quantum criticality one then expects to find branch cuts instead in the optical response, and these should obey generically energy-temperature scaling. However, in the known cases where this is well understood, the exponents are strongly constrained by e.g. the conservation of charge\cite{Lucas:PRB2018}. The conformal tail is somehow of this kind although we are not aware of any explanation that makes sense. 

\section{A brief survey of the experimental optical response.}\label{exp_details}
\begin{figure}
 \includegraphics[width=0.95\columnwidth]{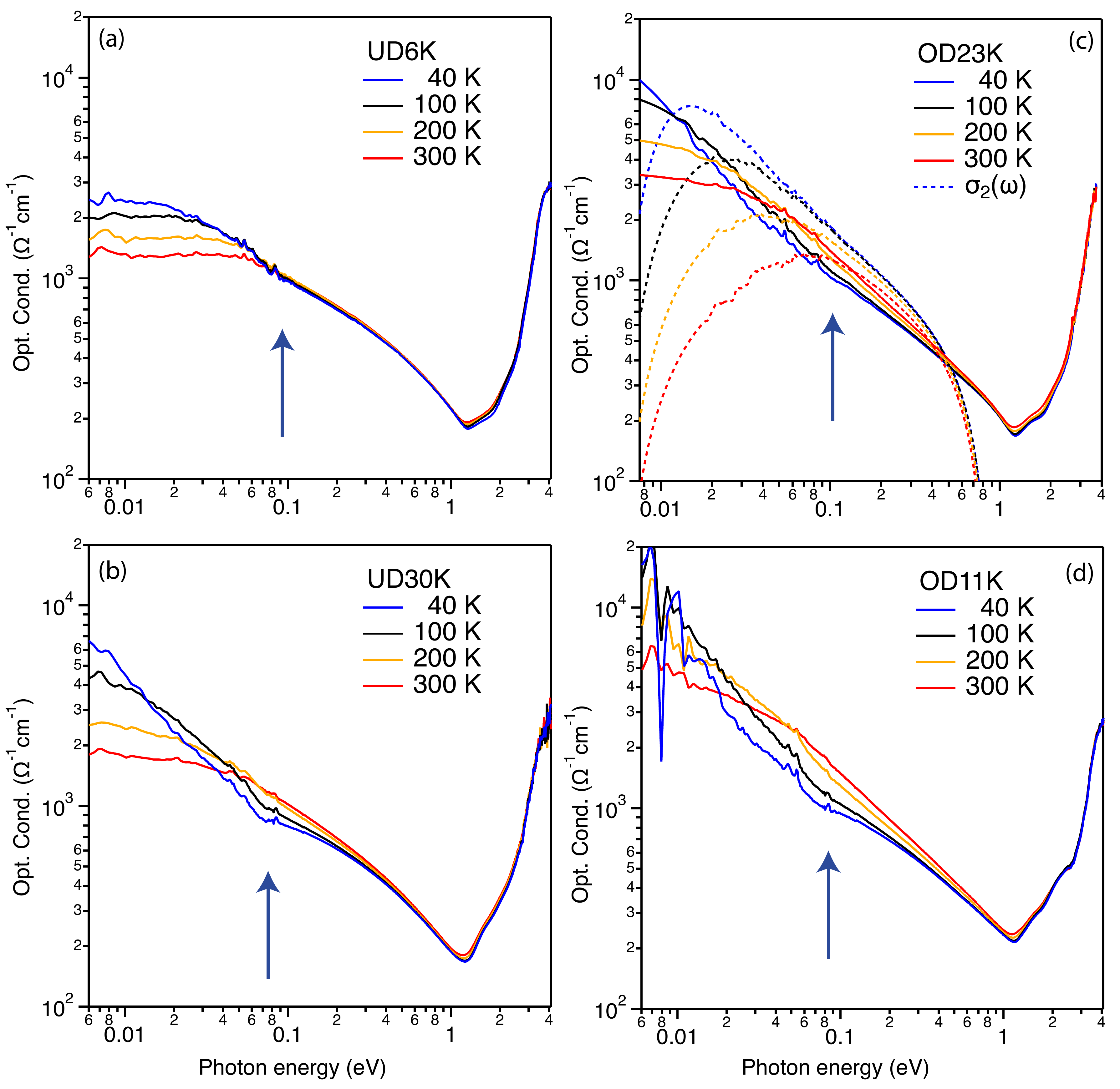}
 \caption{(a-d): Real part of the optical conductivity for a selection of temperatures and doping levels. $\sigma_{1}(\omega)$ at all doping levels and temperatures is characterized by a peak centered at zero frequency and interband transitions above 1 eV. The arrows indicate the energy $\simeq 0.1$ where at low temperature a clear change of slope takes place signalling the presence of two components. Panel (c) also shows the corresponding imaginary part $\sigma_{2}(\omega)$.}
 \label{sigma}
\end{figure}
We focus on the optical properties of BSCO crystals grown using a floating zone method as described elsewhere\cite{Kondo:JESRP2004,Heumen:NJP2009}. The as-grown samples were annealed under oxygen or vacuum conditions to obtain different carrier concentrations. The critical temperature was determined from resistivity measurements and for some samples from ac susceptibility experiments. We use the Presland formula to assign doping levels to our crystals. The validity of this approach has been disputed for BSCO crystals\cite{Lizaire:PRB2021}, but we have recently shown that changes in transport properties and changes in the ARPES spectral functions for the same crystals as measured here are largely consistent with results obtained for other cuprates\cite{Berben:PRM}.
Samples were cleaved immediately before optical experiments were conducted. As described in Appendix \ref{appA}, we use a Kramers-Kronig consistent routine (based on Ref. [\onlinecite{Kuzmenko:RSI2005}]) to transform reflectivity data to the complex optical conductivity, $\sigma_{1}(\omega,T)$. The result is presented in Fig. \ref{sigma}a-d for doping levels spanning the superconducting dome. The optical conductivity of BSCO follows a `hierarchy of energy scales' \cite{Meevasana:PRB2007} where there is a high energy component (the interband response) and two low energy components associated with the valence electrons. In the following, we will start at high energy and comment on each of the three components individually.

The high energy, interband response starts around 1.25\,eV with a weak structure, followed by a stronger transition. In Ref. [\onlinecite{Heumen:NJP2009}], the latter transition was shown to be part of a broader structure with a maximum around 5.5\,eV, based on ellipsometry measurements. The new data presented in Fig. \ref{sigma} is based only on reflectivity experiments and is consistent with these earlier measurements. In what follows, we include the data from Ref. \onlinecite{Heumen:NJP2009} (results reproduced in Appendix \ref{appA}) and find that our observations do not depend on data based on reflectivity only or combined reflectivity and ellipsometry measurements. As shown in Ref. \onlinecite{Heumen:PRB2007} one can use the Clausius - Mossotti relation to estimate the impact of the polarizability of oxygen on the dielectric function:
\begin{equation}\label{Eq:Claussius}
\epsilon_{\infty,IR}\approx 1+\frac{4\pi N\alpha/V}{1-\frac{4\pi}{3}N\alpha/V}= 1+\frac{\alpha_{0}}{1-\gamma\alpha_{0}}
\end{equation}
using the cell parameters for Pb doped BSCO from Ref. [\onlinecite{Chong:physC1997}] and the polarizability of oxygen ($\alpha\,=\,3.88\times 10^{-24}$\,cm$^{-2}$) we estimate that $\epsilon_{\infty,IR}\approx\,4.7\pm0.1$. As explained in appendix \ref{appA}, we use Drude-Lorentz models as a first step in the Kramers-Kronig transformation of the reflectivity data. From these models we obtain an estimate of the contribution of interband transitions to the dielectric function. By summing the contributions of the interband transitions together we find values for $\epsilon_{\infty,exp}$ ranging from 4.2 to 5 with no clear trend. The average value obtained from all samples is $\varepsilon_{\infty,exp}\approx 4.6$, which is in good agreement with the estimate obtained from the Clausius-Mossotti relation. 
\begin{figure}
 \includegraphics[width=0.95\columnwidth]{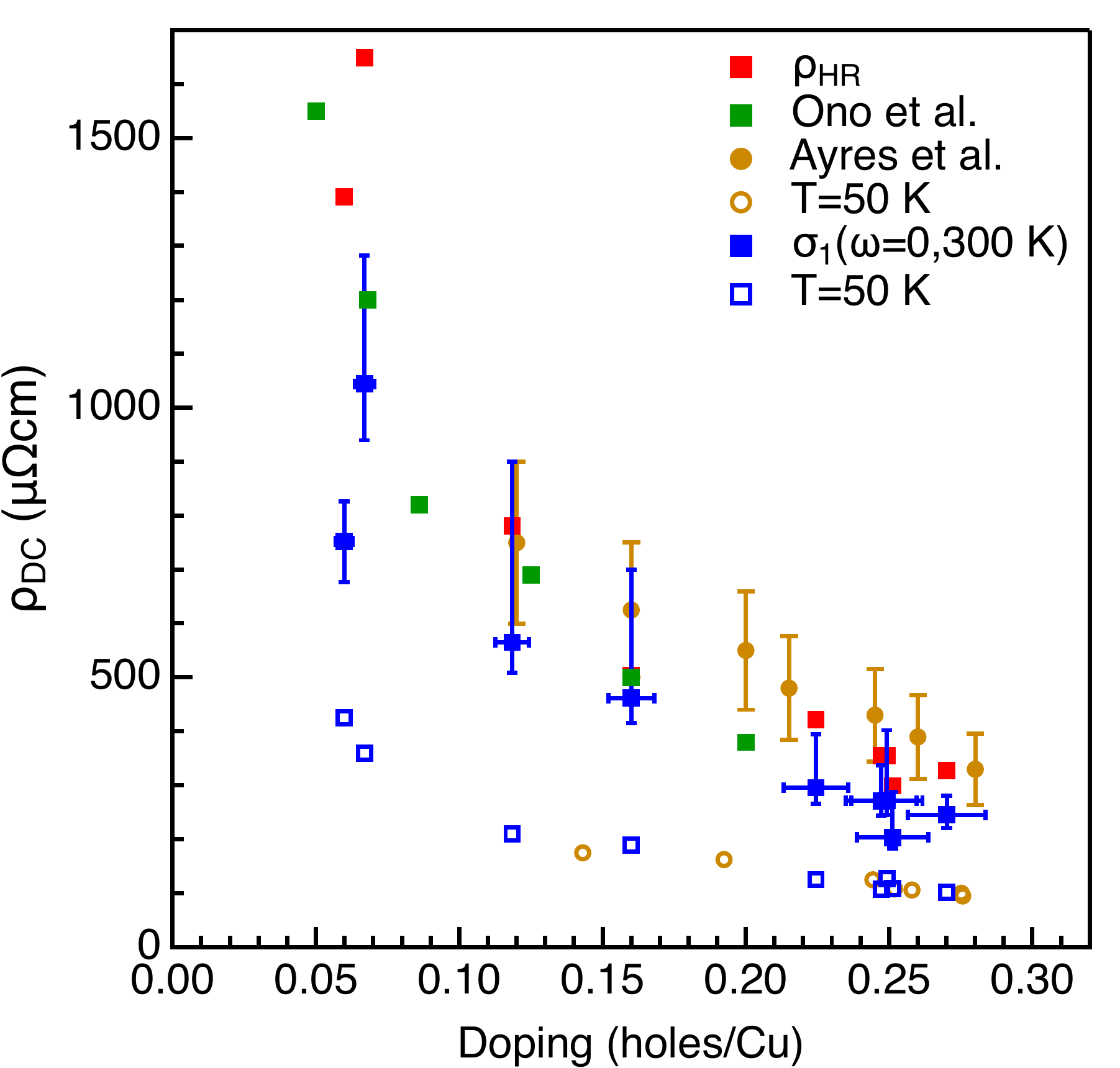}
 \caption{Comparison of DC resistivity values obtained by optical and transport experiments. Red squares indicate the resistivity values obtained from the Hagen-Rubens analysis (see appendix \ref{appA}). Green squares are taken from Ref. \cite{Ono:PRB2003}, while orange circles are measured on similar crystals as used for the optical experiments and are taken from Ref. \cite{Ayres:nature2021}. Blue squares are obtained by extrapolating the optical conductivity to zero frequency. Open symbols indicate the comparison between transport and optics at 50 K.}
 \label{dcrho}
\end{figure}
The main interest of this paper is in the intraband response associated with the valence electrons. At room temperature, $\sigma_{1}(\omega,T)$ below 1 eV is characterized by a single Drude-like peak centred at zero frequency, with an extrapolated DC conductivity that is in excellent agreement with previously published transport data, see Fig. \ref{dcrho}. We note that for all doping levels studied, the room temperature response is manifestly non-Drude in the sense that we never see a $\omega^{-2}$ fall-off. Instead, we find a steadily increasing exponent from -0.5 for UD sample to close to -1 for OD samples. These deviations from classical Drude behaviour have previously given the impetus to describe the data in terms of a generalized Drude response \cite{Dordevic:PRB2005, Schachinger:PRB2006, Hwang:PRB2007, Heumen:PRB2009,Basov:RMP2011}. 
\begin{figure*}[t]
\includegraphics[width=1.95\columnwidth]{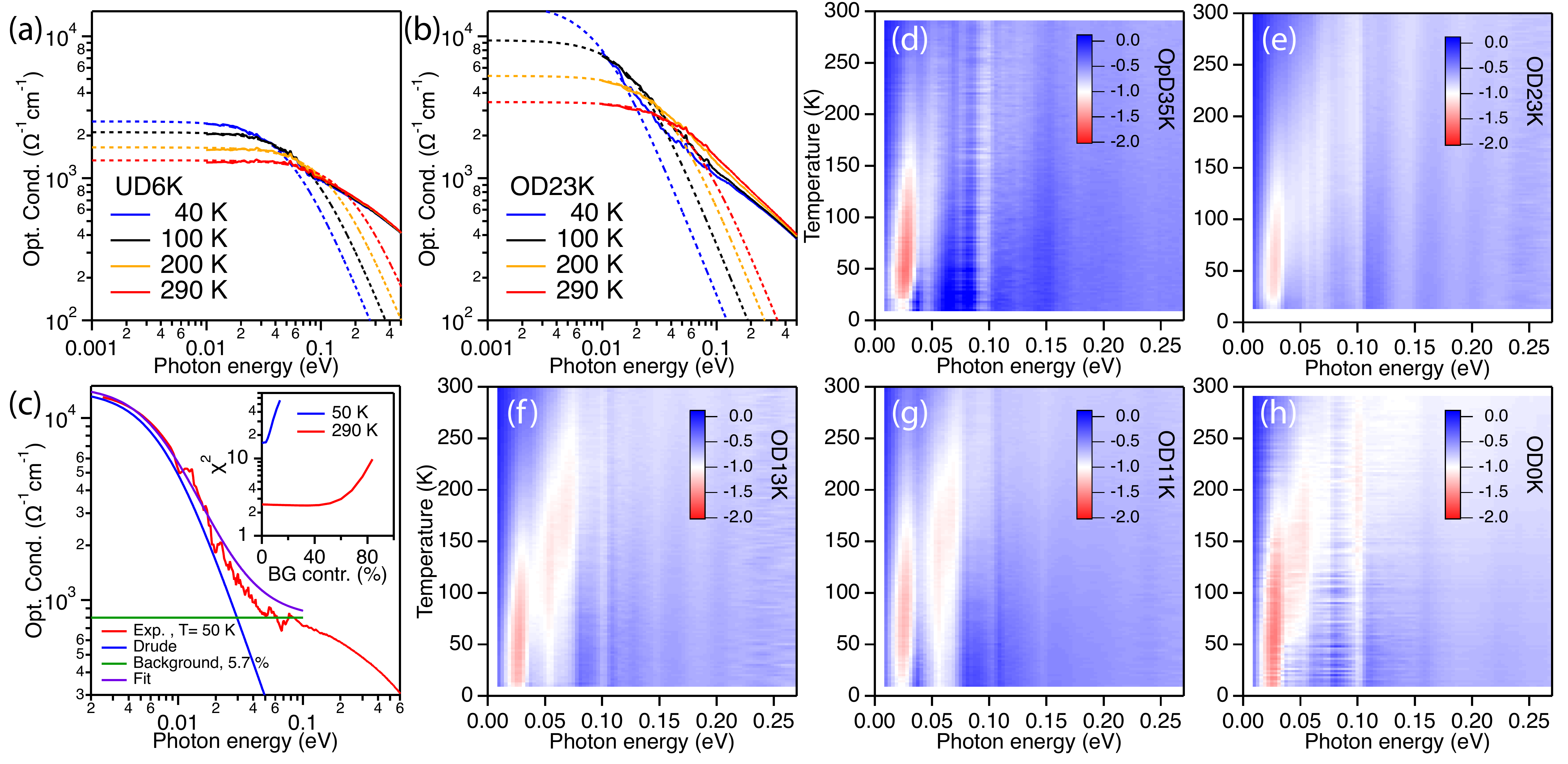}
\caption{(a,b): Real part of the optical conductivity for an UD30K and an OD23K sample, plotted on a log-log scale. Dashed lines are fits with a single Drude response of the low energy conductivity. (c): Estimate of the possible non-Drude contribution at the lowest energies. $\sigma_{1}(\omega)$ is fitted with a sum of a frequency independent background and. Drude peak. The inset shows the quality of the fit, $\chi^{2}$, as function of the percentage of background contribution to the DC conductivity. (d-g) False colour plot of the logarithmic derivative of the real part of the optical conductivity, $\partial\log{\sigma_{1}(\omega)}/\partial\log{\omega}$. The energy and temperature range where a Drude falloff with an exponent of -2 is observed is indicated in red. For underdoped samples no Drude response is observed, while close to optimal doping a narrow Drude response becomes apparent. Vertical lines between 0.02 and 0.07 meV correspond to optical phonons. }
 \label{drudefits}
\end{figure*}
However, as temperature decreases it becomes apparent that the optical conductivity below 1 eV consists of two components. To highlight this, $\sigma_{1}(\omega,T)$ is presented in Fig. \ref{sigma} on a log-log scale. At low temperature, a clear change of slope is visible around 0.1 eV (highlighted by an arrow). At low temperature (40 K, sufficiently far away from $T_{c}$), the slope of the conductivity approaches the Drude slope of -2 and changes around 0.1 eV to approach the same value as seen in the room temperature data. A similar two component response has been inferred previously in bi-layer BSCCO and LSCO \cite{Quijada:PRB1999,Marel:nature2003,Michon:PRR2021}. In section \ref{nofrillsDrude}, it will become clear that these two components do indeed correspond with a genuine Drude response at low energy crossing over to the conformal tail. What is new here is that the low $T_{c}$ of BSCO combined with high quality data makes the two components clearly visible in the raw data without any need for further analysis. We can therefore analyze these two components in detail as function of energy, temperature and carrier concentration. As announced, we will first zoom in on the Drude part, then turn to the conformal taill and finally we will analyze the nature of the crossover between these two very distinct regimes.

\section{The low energy response: the no-frills Drude interpretation}\label{nofrillsDrude}
Let us first zoom in on the low energy regime, below $\omega \simeq 0.1$ eV. As discussed already in relation to Fig. \ref{sigma}, a peak is observed centred at $\omega =0$ that becomes quite sharp at low temperature and which was assigned to a Drude response early on\cite{ishiguro:book1990}. An important question to answer is: to which degree is this a Drude peak? Given the high quality of our optical data, together with their large dynamical range we are in the position to analyze it with precision.

Let's first find out how well the data can be interpreted in terms of the elementary Drude form Eq. (\ref{Simpledrude}), characterized by a frequency independent but temperature dependent momentum relaxation time $\tau_P$. In Section \ref{Matthiessencrossover} we will turn to the possible frequency dependence of the optical self-energy as defined in Eq. (\ref{Memfunction}). There we will find that because of the overlap with the conformal tail around $0.05-0.1$ eV such a frequency dependency cannot be extracted with confidence. The take home message is that within a restricted temperature range the peak can be assigned with high confidence to this elementary Drude response governed by a Planckian-type momentum relaxation rate. 
\begin{figure*}[t]
\includegraphics[width=1.95\columnwidth]{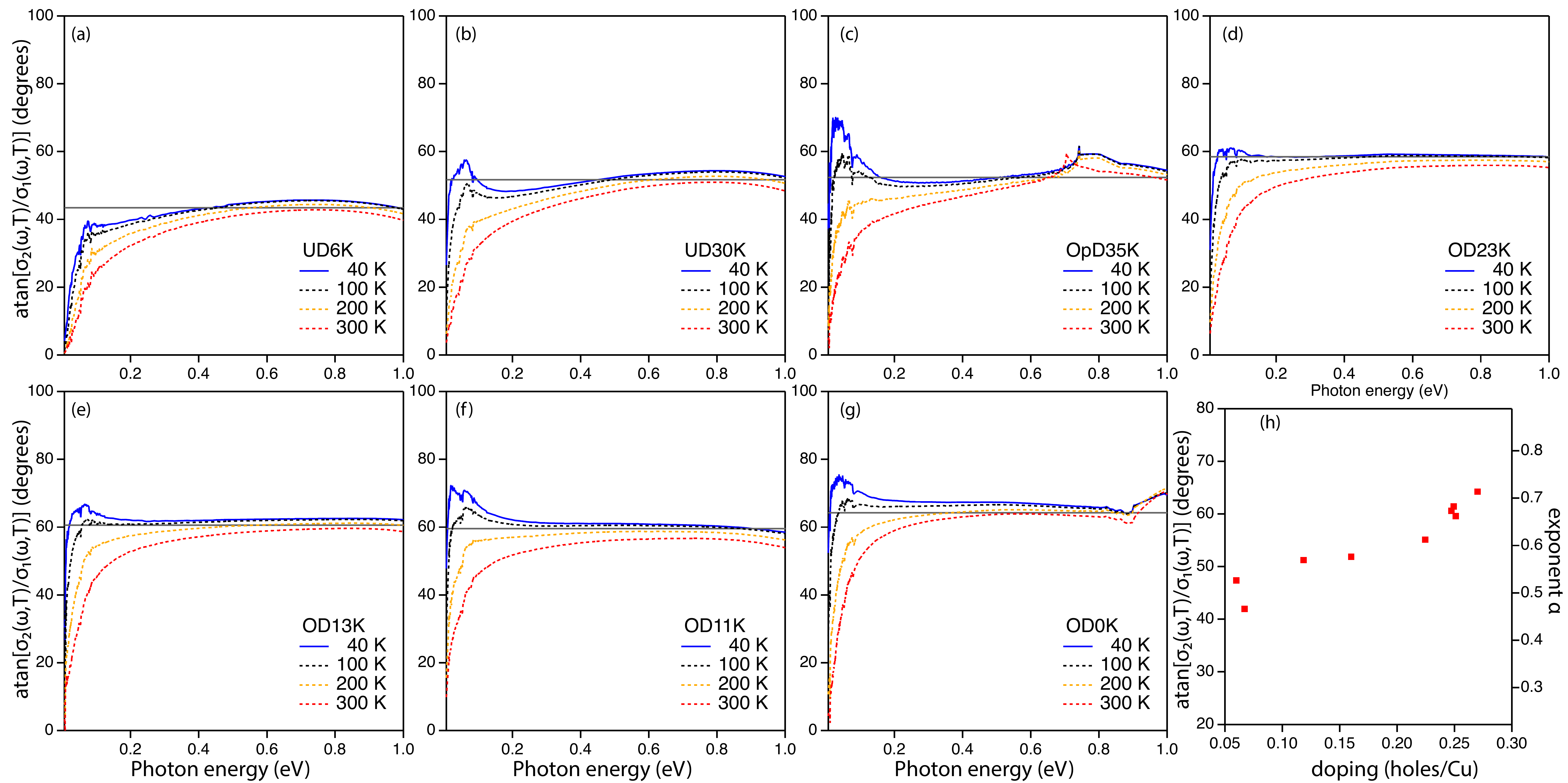}
\caption{(a-g): the phase angle for a selection of doping levels. Below optimal doping, it has a weak frequency dependence, while for higher doping levels the phase angles at low temperature are almost frequency independent above 0.2 eV. (h): doping dependence of the phase angle. }
 \label{phaseangle}
\end{figure*}

The real part $\sigma_1(\omega)$ of this `no-frills' Drude describes a half-Lorentzian. As shown in Fig. \ref{drudefits}a,b, this has a characteristic shape, which in a log-log plot consists of a frequency independent plateau followed by a $\omega^{-2}$ fall off (the dashed lines in the figure; more data presented in appendix \ref{appB}). In order to determine whether such a Drude behaviour is present in the data one should have enough dynamical range to see both at least part of the plateau as well as the $\omega^{-2}$ fall off. As can be seen for the UD6K data presented in Fig. \ref{drudefits}a, and in this case the Drude response is always broad and the $\omega^{-2}$ falloff is masked by the conformal tail. However, this criterion is full-filled by the experimental data for optimally to overdoped samples. Although we do not have data below $\sim 5$ meV, the extrapolated DC values are consistent at high and low temperature with the available transport data (see Fig. \ref{dcrho}) leaving no doubt that the plateau is present. The data also reveal the $\omega^{-2}$ fall-off over a small range of photon energies at lower temperatures. This figure reveals that for energies above approximately 40 meV, the frequency dependence changes and a slower fall off takes over, signalling the cross-over to the conformal tail.

To test the robustness with which we can assign the response below 0.1 eV to a Drude peak, we model the experimental data as a combination of a Drude response and a background contribution, see Fig. \ref{drudefits}c. This background would be interpreted as the low energy extrapolation of the mid-infrared response, which in principle could extend to zero frequency. We use a standard least-square measure, $\chi^{2}=\sum\sqrt{\sigma_{1}(\omega_{i})-f(\omega_{p},\Gamma,\sigma_{BG})}$, to determine the closeness of the fit to the experimental data as we increase the background value of the optical conductivity. The example shown in Fig. \ref{drudefits}c exhibits the Drude component in blue, while the background is shown in green. The inset shows the calculated $\chi^{2}$ as function of the background contribution expressed in percentage of the total DC conductivity. We observe that at low temperature the quality of the fit quickly deteriorates when we add a background. At room temperature, the quality of the fit remains independent of the background contribution up to 50 $\%$. From this analysis we can conclude (i): as temperature approaches $T_c$ the low energy Drude response constitutes more than 90 $\%$ of the total conductivity below 0.1 eV and (ii): at higher temperatures we loose the sensitivity to determine whether a Drude response is operative. 

For energies larger than the current relaxation rate, the Drude response becomes approximately:
\begin{equation}
\lim_{\omega\gg\Gamma}\sigma_{1}(\omega)\approx \frac{\omega_{p}^{2}}{\omega^{2}}
\end{equation}
From this it follows that the logarithmic derivative, $\partial\log{\sigma_{1}(\omega)}/\partial\log{\omega}$ should approach the value of -2 for a Drude response. Figure \ref{drudefits}d-g shows this logarithmic derivative in a false colour scale. The scale is chosen such that an exponent of -2 has a red color. For the OpD35K sample, the Drude response is clearly visible at low temperature. As temperature increases, the $\omega^{-2}$ response moves to higher energy, corresponding with the increasing relaxation rate. Around 150 K, the Drude response has broadened to the point where it starts to merge with the incoherent high energy response. At high temperature and low energy, the exponent approaches zero which is consistent with a plateau in the optical conductivity (see Fig. \ref{drudefits}a,b).

\section{Tracking the branch cut.}\label{branchcut}
As we argued in the above the fingerprint of the branch cut response Eq. (\ref{Eq:branchcut}) is an asset for the analysis of the data. As we will see in the next section, there is an ambiguity with regard to the precise nature of the crossover from the Drude- to conformal tail regimes. However one can invariably identify an energy scale below which the conformal tail is suppressed. Although the origin of the conformal tail is presently completely in the dark its gross properties may be best understood as reflecting some form of {\em bound} optical response -- it may be viewed as the analogue of interband transitions in the strongly interacting electron soup. 

The first task is to find out whether the conformal tail fingerprints (power-law conductivity and constant phase angle) characterize the optical conductivity of our single layer BSCO samples. In the bilayer BSCCO system this was previously investigated, indicating that the conformal tail appears to be present at all doping levels with varying exponent \cite{Marel:nature2003,Hwang:JPCM2007}. 

In Fig. \ref{phaseangle}a-g we show an overview of the phase angle for our single layer BSCO samples. As in the bilayer system, we find that in the range $\sim 0.2 - 0.8$ eV both $\sigma_{1} (\omega)$ and $\sigma_{2} (\omega)$ are characterized by an algebraic fall-off, as well as a near frequency independent phase angle that is consistent with the exponent determined from the algebraic fall-off (indicated by the grey line). Similar to the two previous works, we observe a small deviation from a perfect frequency independent phase below optimal doping. We also observe that at elevated temperature the phase angle becomes more curved. Importantly, the branch cut tail persists over the whole doping range and is well discernible in even the most strongly overdoped samples. 

This is adding impetus to the notion that the highly overdoped cuprate metals are far from being just weakly interacting Fermi-liquids. Intriguingly, we do resolve yet another notable doping dependence: as shown in Fig. \ref{phaseangle}h, there appears to be a rather sizable dependence of the "anomalous scaling dimension" (exponent) $\alpha$ characterizing the branch cut as function of doping in the overdoped regime. Being rather doping independent in the underdoped regime, we find that it increases roughly proportional to the increasing carrier density upon entering the overdoped regime. We do find that $\alpha_{\sigma} \simeq 0.57$ close to optimal doping, somewhat smaller than what is found in the two layer BSSCO ($\simeq 0.7$). 

When the conformal tail is originating in some form of quantum critical behaviour one expects for very general reasons that this should be governed by energy-temperature ($\hbar\omega/2\pi k_{B}T$) scaling. Based on Eq. (\ref{Eq:branchcut}), one expects that energy and temperature appear on equal footing following a quadrature form. This should result in a scaling collapse of the following form,
\begin{equation}\label{scalingfunc}
\left[(2\pi k_{B}T)^{\alpha}\sigma_{1}(\omega)\right]^{-1}=\left(\frac{\hbar\omega}{2\pi k_{B}T}\right)^{\alpha}
\end{equation}
The result is shown for an overdoped and underdoped sample in Fig. \ref{phasescaling}a,b (other doping levels are shown in Appendix \ref{appB}). At first glance, the temperature exponent derived in this way agrees reasonably well with frequency exponent extracted from the phase angle of Fig. \ref{phaseangle}. However, the deviations below optimal doping are significant and we notice that the scaling collapse is far from perfect in the energy range where we expect it to work best. We will see in the next section that this arises from an additional energy scale in the problem.
\begin{figure}
\includegraphics[width=0.99\columnwidth]{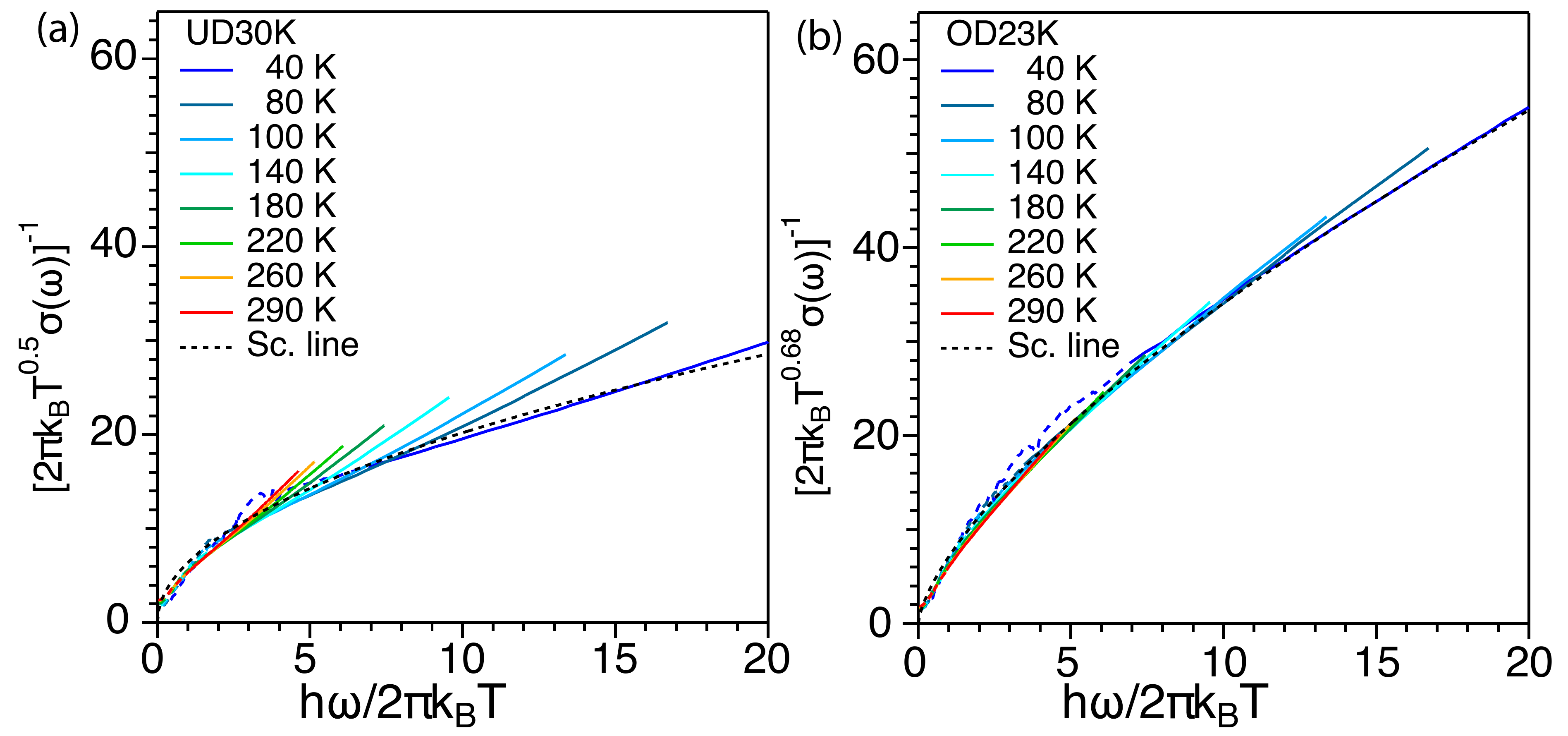}
\caption{(a,b): scaling of the real part of the optical conductivity for an underdoped and an overdoped sample. The dashed line is the calculated scaling form based on the temperature exponent. }
 \label{phasescaling}
\end{figure}

\section{The crossover regime: are the conductors in series or parallel ?} 
\label{crossoverreg}

Arrived at this point we have confirmed the case presented in Ref. \cite{Marel:nature2003,Marel:AP2006} that the low frequency optical response ($\omega < 0.05$ eV) of single layer BSCO is captured by a simple Drude response characterized by a seemingly frequency independent momentum relaxation time while the high energy response ($\omega > 0.1$ eV) carries the tell tale signs of an `incoherent' conformal tail response implying a complete decoupling of the electrical currents from momentum. But what happens in the middle, how does the crossover between these two distinct responses work? 

The crucial observation is that there {\em has} to be an energy scale associated with the onset of the incoherent response. Extrapolating $| \omega |^{-\alpha_{\sigma}}$ to $\omega \rightarrow 0$ would imply a divergent response while instead we find that the Drude response completely dominates. Given that there is no clear signature of this crossover, we have to resort to modelling. 

A crucial question in this regard is, do the Drude- and incoherent parts behave like conductors in parallel (`Anti-Matthiessen', AM) or in series (`Matthiessen', M)? As we already emphasized, the first possibility seems to make more physical sense. Metaphorically, it is like the division of the free carrier response (the Drude part) and the `bound' interband excitations in conventional systems where the latter are by default decoupled from total momentum. This implies that one has to add the {\em conductivities} of the two sub-systems. This is the antithesis of what is often assumed in transport experiments. Asserting that the temperature dependence of the resistivity is rooted in the sum of different relaxational contributions, these different channels add in series and for the optical response this would manifest itself in a memory function type behaviour, Eq. (\ref{Memfunction}). 

In section \ref{Matthiessencrossover} we will take up the analysis of the crossover in this memory function guise, but let us first see how far we can get with the AM interpretation of the optical response. 

\subsection{The Anti-Mathiessen interpretation.}\label{AntiMatcrossover}
\begin{figure}
\includegraphics[width=1\columnwidth]{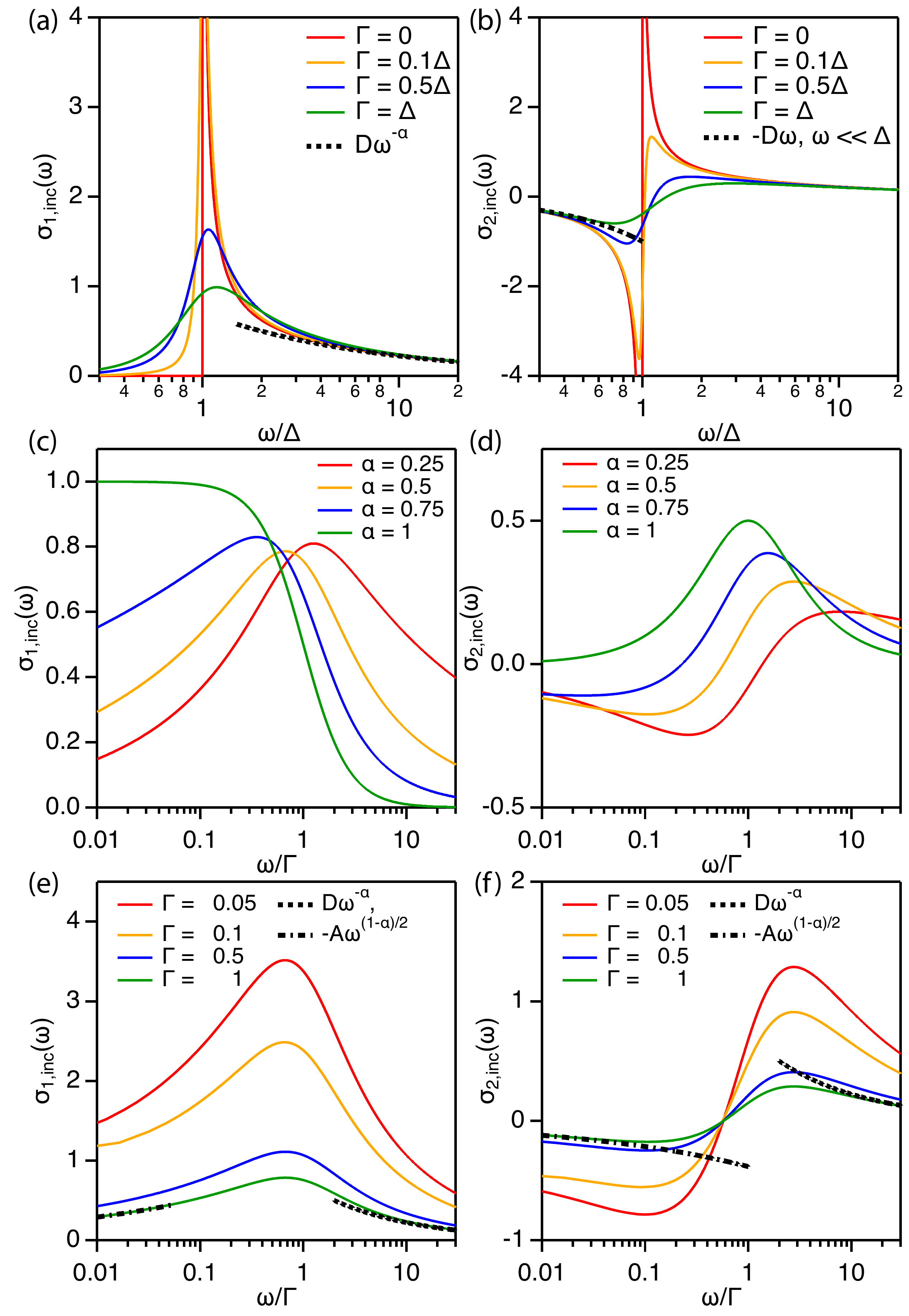}
\caption{(a,b): Real (a) and imaginary (b) components of $\sigma_{inc}(\omega)$ for $\Delta\neq\,0$ and increasing $\Gamma$. For $\Gamma\,=\,0$, $\sigma_{1}(\omega)$ vanishes below $\Delta$ and has a divergence $\propto\varepsilon^{-(\alpha+1)/2}$ at $\Delta$. As $\Gamma$ increases, the divergenze is smoothed out and shifts away from $\Delta$. At high energy we recover the conformal tail as indicated by the dashed line. (c,d): $\hat{\sigma}^{inc}(\omega)$ for $\Delta\,=\,0$ and fixed $\Gamma$ for a range of experimentally relevant exponents. For $\alpha\,=\,1$ we retrieve a Drude response. As $\alpha$ decreases a finite frequency peak develops. (e,f): same but now for $\Delta\,=\,0$ and increasing $\Gamma$. The maximum conductivity decreases with increasing $\Gamma$. Note that there are now two regimes where the conductivity follows a powerlaw (dashed and dash-dotted lines), but with two different exponents. In panels (a-d) we used $\alpha\,=\,0.5$. 
}
\label{modLorentz}
\end{figure}

In the anti-Mathiessen interpretation we assume two contributions to the optical conductivity and use the following model, 
\begin{eqnarray}
\hat{\sigma} (\omega) & = & \hat{\sigma}^D (\omega) + \hat{\sigma}^{inc.} (\omega) \nonumber \\
\hat{\sigma}^D (\omega) & = & \frac{D_{Dr}}{4 \pi} \frac{1}{ \Gamma_{Dr} + i\omega } \nonumber \\
\hat{\sigma}^{inc.} (\omega) & = & \frac{-iD_{inc}\omega}{\left(\Delta^2 - \omega^{2}-i \Gamma_{inc}\omega\right)^{\beta}}
\label{AntiMatfit}
\end{eqnarray}
for the complex optical response with $D_{Dr},D_{inc}$ referring to the the Drude and incoherent spectral weight respectively. $\Gamma_{Dr}$ is the Drude width. $\hat{\sigma}^{inc.} (\omega)$ is inspired by a simple Lorentz oscillator at energy $\Delta$ and with damping $\Gamma_{inc}$: when $\beta =1$ this form is precisely that of a Lorentz oscillator. We now just assert that its `engineering' dimension turns anomalous, $\beta\neq\,1$. For $\omega >> \Delta, \Gamma_{inc}$ the branch cut, Eq. \ref{Eq:branchcut}, is recovered. Comparing to Eq. \ref{Eq:branchcut}, we find the scaling dimension $\beta=(\alpha+1)/2$ with the conformal tail exponent $\alpha$. Thus, $\hat{\sigma}^{inc.} (\omega)$ is no more than a phenomenological way to wire in in a flexible way a branch-cut that is disappearing below a characteristic scale. At high energy it is characterized by a `relevant' exponent $\alpha > 0$ such that $\sigma_1$ diverges for $\omega \rightarrow 0$. This is interrupted at some energy scale below which the low energy spectral weight is suppressed. Importantly, this is done in a Kramers-Kronig consistent form and we will use this powerful constraint on the data to our advantage. 

$\Gamma_{inc} = 0$ encodes for a `hard gap' in $\hat{\sigma}^{inc.} (\omega)$. The real part of $\hat{\sigma}^{inc.} (\omega)$ vanishes for $\omega < \Delta$ to jump up discontinuously at $\Delta$, with a divergence that goes as $1/ \varepsilon^{(\alpha +1)/2}$ at $\omega = \Delta + \varepsilon$. This merges smoothly into the conformal tail when $\omega$ becomes large (see the case $\Gamma\,=\,0$ in Fig. \ref{modLorentz}a). The effect of a $\Gamma_{inc} < \Delta$ is to just smear this hard gap over a scale $\sim \Gamma$ as Fig. \ref{modLorentz}a shows. 

There is however yet a different way of manipulating the low energy spectral weight. Note that we obtain a Drude form by setting $\Delta = 0$ and $\beta\,=\,1$, corresponding to the green line in Fig. \ref{modLorentz}c. What happens when the scaling dimension $\beta$ turns anomalous? Fig. \ref{modLorentz}c show that for decreasing $\alpha$ the Drude response is `pushed' away from $\omega\,=\,0$. From Eq. \ref{AntiMatfit} we find that the incoherent conductivity simplifies to, 
\begin{equation}
 \hat{\sigma}^{inc.} (\omega) = \frac{D_{inc}} {( i \omega)^{\alpha} } \; \times \frac{1}{ ( 1 + i ( \Gamma_{inc} / \omega ))^{(\alpha +1)/2} }
\label{Gammaincdamped}
\end{equation}
we read off that for $\omega >> \Gamma_{inc}$ we recover the conformal tail. However, zooming in on the low energy regime $\omega << \Gamma_{inc}$ this becomes, 
\begin{equation}
\lim_{\omega \rightarrow 0} \hat{\sigma}^{inc.} (\omega) = \frac{D_{inc}}{\Gamma_{inc}^{(\alpha+1)/2} } \frac{1}{ ( i\omega)^{(\alpha - 1)/2}}
\label{Gammaincdamplow}
\end{equation}
We see that this describes yet another branch cut but with an altered scaling dimension. For $\alpha < 1$ this turns irrelevant, with the spectral weight in $\sigma_1$ decreasing to zero in a branch-cut fashion when $\omega \rightarrow 0$. This models the flow from an `UV' (high energy) CFT to an `IR' (low energy) CFT characterized by different scaling dimensions, capturing a smooth crossover from the different scaling behaviour at high- and low energy. This is illustrated in Fig \ref{modLorentz}c for a representative $\alpha = 0.5$ and varying $\Gamma_{inc}$. The crossover at $\omega \simeq \Gamma_{inc}$ manifests itself as a peak in the real part of $\hat{\sigma}^{inc}(\omega)$.
\begin{figure}
\includegraphics[width=1\columnwidth]{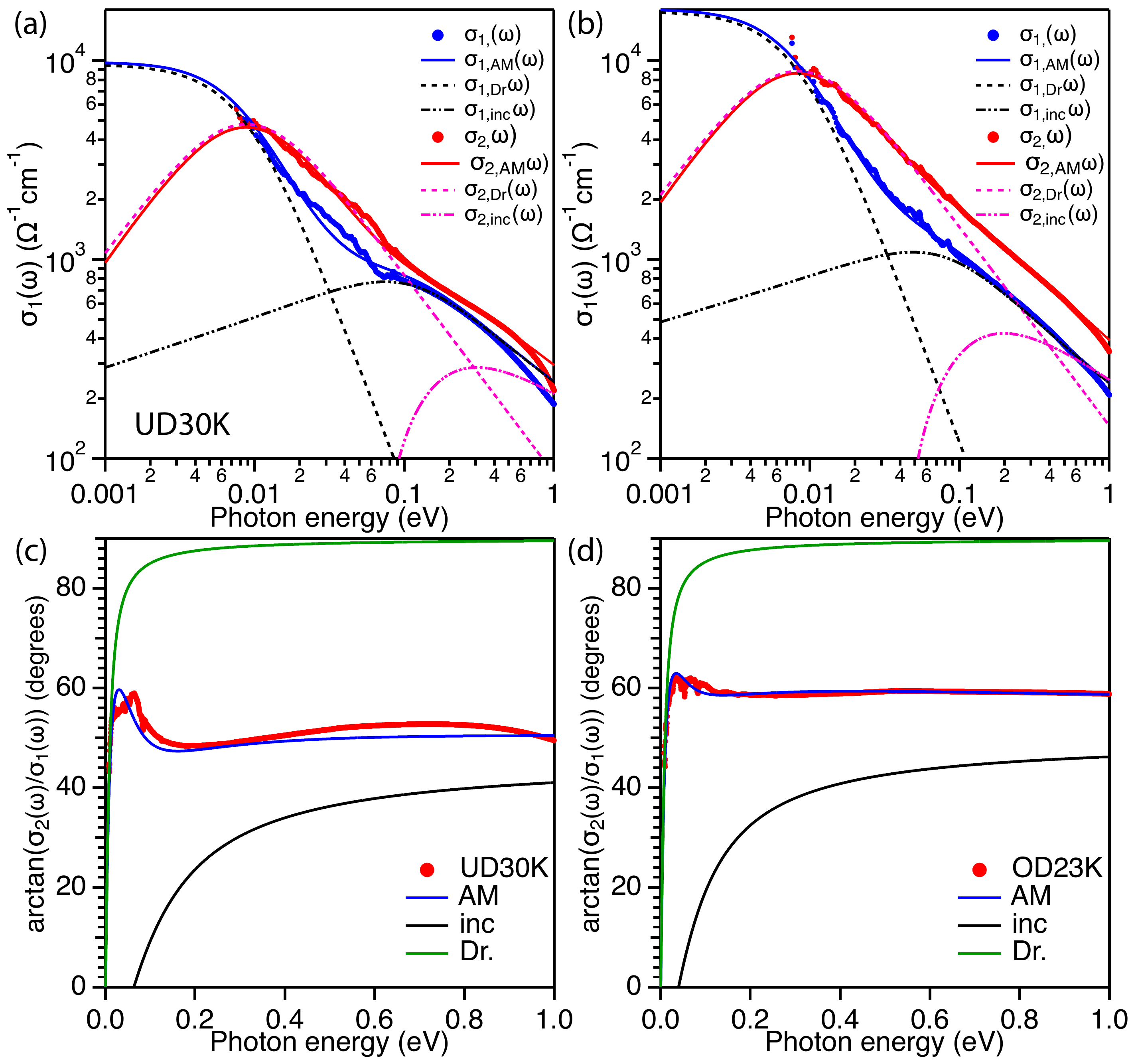}
\caption{(a,b): Decomposition of $\sigma_{1}(\omega)$ and $\sigma_{2}(\omega)$ of the UD30K and OD23K samples at 40 K in a Drude response (dashed) and conformal tail (double-dashed). (c,d): corresponding experimental phase angle compared to the phase angle of the two individual components and the anti-Mathiessen result. To obtain these fits we have taken $\Delta\,=\,0$ in both cases (see text for details).}
\label{AMdecomp}
\end{figure}
Such details of $\hat{\sigma}^{inc.} (\omega)$ in the low energy regime are completely shrouded from $\sigma_1(\omega)$ given the domination of the Drude part. From $\sigma_{1}(\omega)$ alone it is difficult to get any detailed information on what to take for $\Delta$ and $\Gamma_{inc}$ beyond a crude estimate for the damping of the gap -- an infinitely hard gap would give rise to a discontinuity in the total $\sigma_1(\omega)$ that is not present. However, this changes drastically considering the imaginary part of the conductivity $\sigma_2(\omega)$. In Fig. \ref{modLorentz}b,d,f, we present the $\sigma_2(\omega)$ corresponding to the previously discussed representative cases. When the `hard' gap dominates ($ \Delta > \Gamma_{inc}$) one finds the typical large `wiggle' in $\sigma_2(\omega)$ imposed by Kramers-Kronig consistency. Similar to $\sigma_{1}(\omega)$, the divergent response is smoothed out as $\Gamma_{inc}$ increases. As we will see next, the combination of real and imaginary components of the optical response will allow us to make concrete statements about the ratio $\Gamma_{inc}/\Delta$.

To illustrate the fitting, we show in Fig. \ref{AMdecomp} the details of the analysis for an underdoped and overdoped example. Panels \ref{AMdecomp}a,b compare the fit (solid lines) using the two component model of Eq. (\ref{AntiMatfit}) with the experimental optical conductivity (symbols). Also indicated are the individual components (dashed and double dashed lines).
\begin{figure}
\includegraphics[width=0.95\columnwidth]{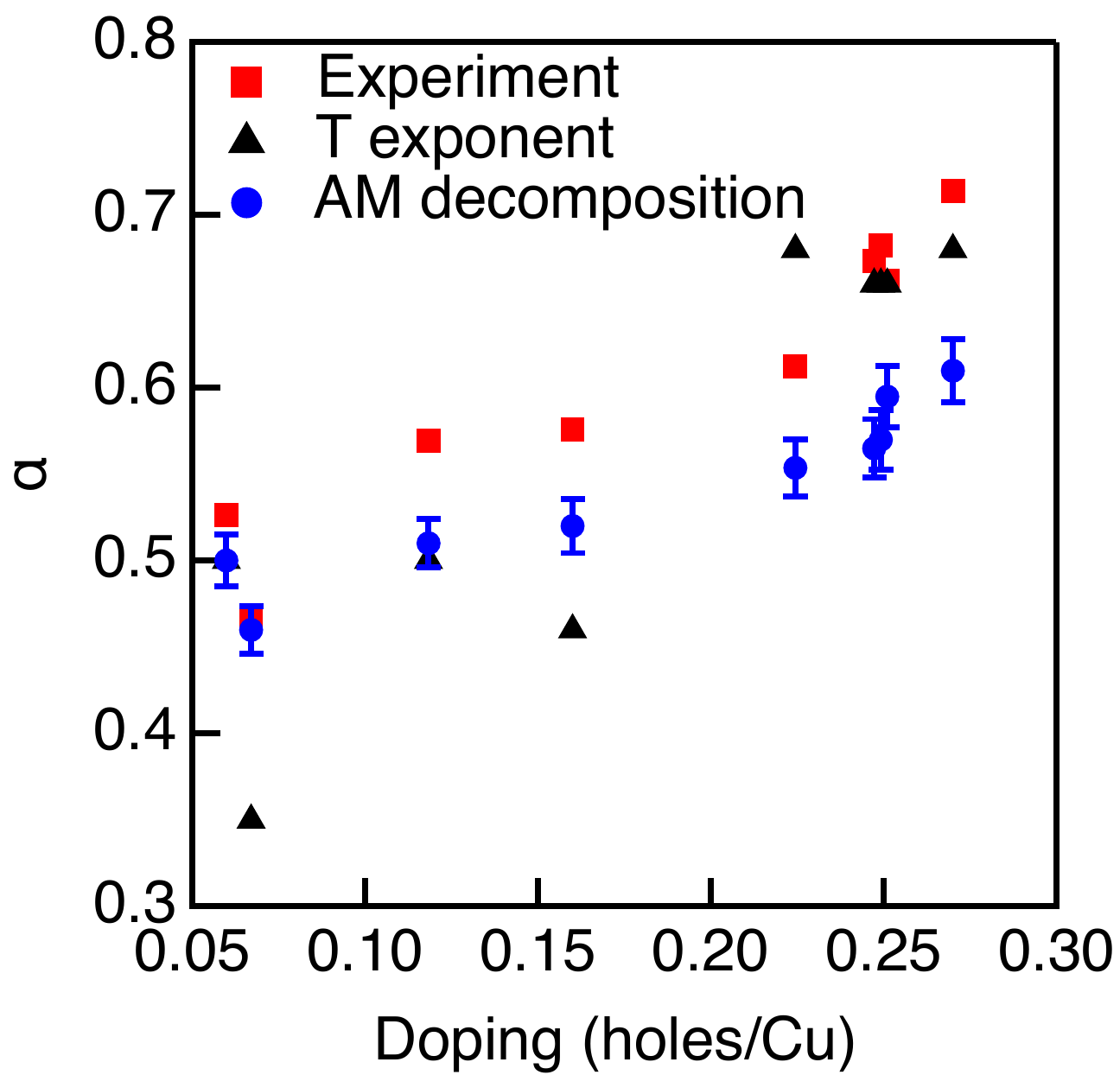}
\caption{Conformal tail exponent as function of doping from the anti-Mathiessen decomposition (blue circles). The red squares are the values obtained directly from experiment (Fig. \ref{phaseangle}h), while the black triangles are obtained from the scaling analysis of section \ref{branchcut}.}
\label{AMparams}
\end{figure}
In all data presented here we have assumed $\Delta\,=\,0$, but we can obtain reasonable fits for $\Delta\,<$\,25 meV, as long as $\Gamma_{inc}>3\Delta$. To obtain this bound requires the use of both real and imaginary components of the optical response. For smaller values of $\Gamma_{inc}$, the strong divergencies in Eq. \ref{AntiMatfit} give rise to structures not observed in the experimental optical conductivity. This is best seen by looking at the phase angle (Fig. \ref{AMdecomp}c,d). The blue line shows the best fit to experimental result (red circles). The experimental `peak-dip' structure in the phase angle around 0.1 eV, is directly related to the conformal tail and controlled by the ratio of $\Gamma_{inc}/\Delta$. To reproduce the experimental phase angle requires $\Gamma_{inc}/\Delta\,>\,3$. For smaller ratios of $\Gamma_{inc}/\Delta$ and/or values of $\Delta\,>\,25$ meV, the peak-dip structure in the phase angle becomes more and more violent and we lose the agreement with the experimental data. Comparing the two sample, we find a much more pronounced peak-dip structure at low energy for the UD30K sample, which we trace back to a significantly larger $\Gamma_{inc}$ ($\Gamma_{inc}\approx$\,115\,meV for UD30K versus $\Gamma_{inc}\approx$\,80\,meV for OD23K).

Despite the importance of the conformal tail response to reproduce the data in the crossover region, Fig. (\ref{AMdecomp})a,b emphasizes that the low energy spectral weight is overwhelmingly that of the Drude component, leaving intact our conclusion from the previous section. The possibility that there is an `IR CFT' type response at work associated with the incoherent response should not to be taken too seriously. The simple form Eq. (\ref{Gammaincdamped}) is just a convenience to get $\sigma_1(\omega)$ and $\sigma_2(\omega)$ in balance in the {\em high} energy regime and we have no argument rendering the extrapolation at low frequencies to be of this simple kind. It is very well possible that at low energy the response is modified, leaving room for a finite DC contribution\cite{Ayres:nature2021}. On the basis of only the data it is just impossible to specify how it behaves at low frequency. But it does stress that the constancy of the high energy phase angle setting in rapidly above the cross over scale requires quite special analytical properties of this response.

Taking this for granted, this analysis brings three interesting and significant points to the front that appear to be rather independent of these ambiguities.
In the first place, the energy scale associated with $\Gamma_{inc}$ is at least of order 70 - 80 meV. This is essentially the Debye frequency of the cuprates, where the highest phonon mode is the Cu-O stretching mode visible in the the data presented in Fig. \ref{fig:refl}a. We find that this is the minimum $\Gamma_{inc}$ needed for overdoped samples. For optimally doped and underdoped samples, $\Gamma_{inc}$ increases and the largest value found $\Gamma_{inc}\approx$\,0.2 for the UD10K sample.

The second interesting point is summarized in Fig. \ref{AMparams}: the actual conformal tail exponent can be quite a bit smaller than what one would expect based on the experimental data. We find that the observed phase angle is set by the conformal tail exponent and the ratio of the spectral weights $D_{Dr}$ and $D_{inc}$. Based on our analysis, we find that the exponent at optimal doping is close to $\alpha\approx\,0.5$, which is similar to the power law exponents observed in the self energy of angle resolved photoemission experiments\cite{Reber:ncomm2019,Smit:arxiv2021}. Also the doping dependence agrees well with the observed dependence in these experiments.

The third point concerns the temperature dependence of the optical data. For all samples studied, the conformal tail parameters are within our experimental resolution temperature independent. To accurately reproduce the temperature dependence requires optimization of the Drude component, and thus $D_{dr}$ and $\Gamma_{Dr}$ only. We will discuss this temperature dependence in the next sections.

\subsection{The Drude width: the near Planckian relaxation rate.}
\begin{figure}
\includegraphics[width=0.95\columnwidth]{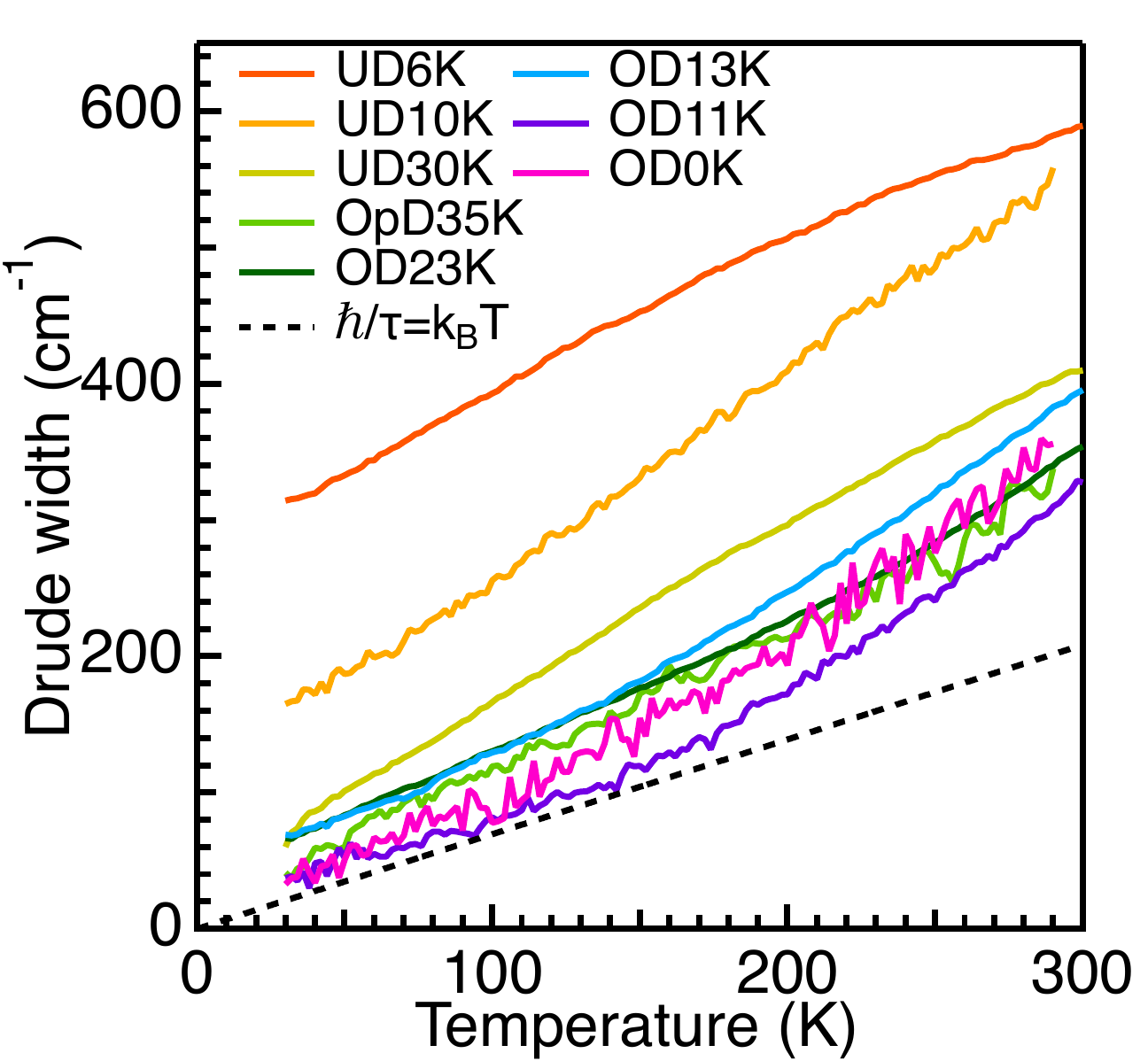}
\caption{Temperature dependence of $\Gamma_{Dr}$ for different doping levels. At high doping the data approaches the Planck limit (dashed line). Below optimal doping the curvature of the scattering rate becomes convex.}
\label{Drwidth}
\end{figure}
Given that the signature of the Drude peak in the form of the $\omega^2$ fall off is no longer discernible at temperatures $\ge 150$ K is no longer discernible (Fig. \ref{drudefits}) it can surely not be claimed that at higher temperatures we are still dealing with a simple single relaxation time Drude response. However, by using Eq. (\ref{AntiMatfit}) we just assert this to be the case and it suffices to quantify the data with fits that are especially of high quality in the low energy regime at all temperatures and dopings. Once again, we assume a single relaxation rate, ignoring a possible energy dependence of the optical self energy that should manifest itself as a deviation of the Drude line form at higher energies. We will see that according to the Mathiessen fitting procedure where we assert that this is entirely due to such a frequency dependency it is a small correction at low energy. In the AM fitting this is completely masked by the `spill over' of the incoherent spectral weight. 

Given these caveats, we can address the question: assuming that the DC resistivity is given by the simple Drude form $\rho = \Gamma_{Dr}/D_{Dr}$ (as is often done) what is the origin of the temperature dependence of $\rho$? As we will discuss next, the Drude weight can be accurately determined and we find it to be  weakly temperature dependent. Hence, the temperature dependence is nearly entirely due to the current relaxation rate $\Gamma_{Dr} = 1 /\tau_J$. 

The results for $\Gamma_{Dr}$ are show in Fig. \ref{Drwidth}. We find that this closely tracks the temperature dependence of the DC resistivity\cite{Ono:PRB2003,Ayres:nature2021}. Ignoring the (strongly) underdoped regime we find that it becomes nearly linear in temperature at optimal doping. It is well known that even at optimal doping the resistivity in the single layer BSCO has a small but discernible curvature even at optimal doping. Upon increasing doping this curvature is getting more pronounced and this is reflected in the behaviour of $\Gamma_{Dr}$. 

As a reference we also show the Planckian relaxation rate (dashed line). The residual resistivity is comparatively large at optimal doping in this family but we see that the slope of $\Gamma_{Dr}$ is here very close to the Planckian value. Perhaps more surprising, the main effect of overdoping is that the residual resistivity is decreasing. Ignoring the (small) curvature, the overall magnitude of the slope continues to be very close to the Planckian value up to the highest doping levels. Relying on the Planckian dissipation as the main signature of strange metals, also in this regard the (strongly) overdoped cuprate metals are not at all different from the optimally doped ones.

\subsection{The spectral weight distributions: the smooth evolution of the Drude weight.}\label{spectralweight}

To complete the connection to transport experiments, we turn to the doping and temperature dependence of the optical spectral weight. The f-sum rule states that the free charge spectral weight is obtained by integrating the measured optical conductivity,
\begin{equation}
D_f = \int_0^{\omega_{c} } d \omega \sigma_1 (\omega)
\label{fsumrule}
\end{equation}

For the cuprates, the cut-off frequency $\omega_{c}$ is often chosen to be of order 1\,eV to separate the `free charge' from `interband transitions'. The quotation marks are a caution: one should not interpret these words in terms of the (usual) non interacting band structure language. The notion of `bound' excitations associated with interband transitions survives in the strongly interacting context dealing with transitions that are in the band structure between completely occupied- and unoccupied bands, setting in around 1 eV. However, the free charge now refers to the physics ruled by doping the Mott insulator which is by itself eventually revolving around the interplay of the Umklapp potential and dominating interactions, that is in all likelihood of the origin of the mysterious conformal tail. 
 
Making use of the unit cell volume, the optical spectral weight can be converted to an effective number of free charge carriers per Cu-O plaquette ($N_{eff}/m^{*}$). This optically determined charge carrier density is unambiguous dealing with a Drude response. Next to its intrinsic importance, it can also be used to test claims regarding the carrier densities as follow from interpreting Hall experiments\cite{Badoux:Nature2016,Putzke:nphys2021,Ayres:nature2021,Lizaire:PRB2021}. 

\begin{figure}
\includegraphics[width=0.95\columnwidth]{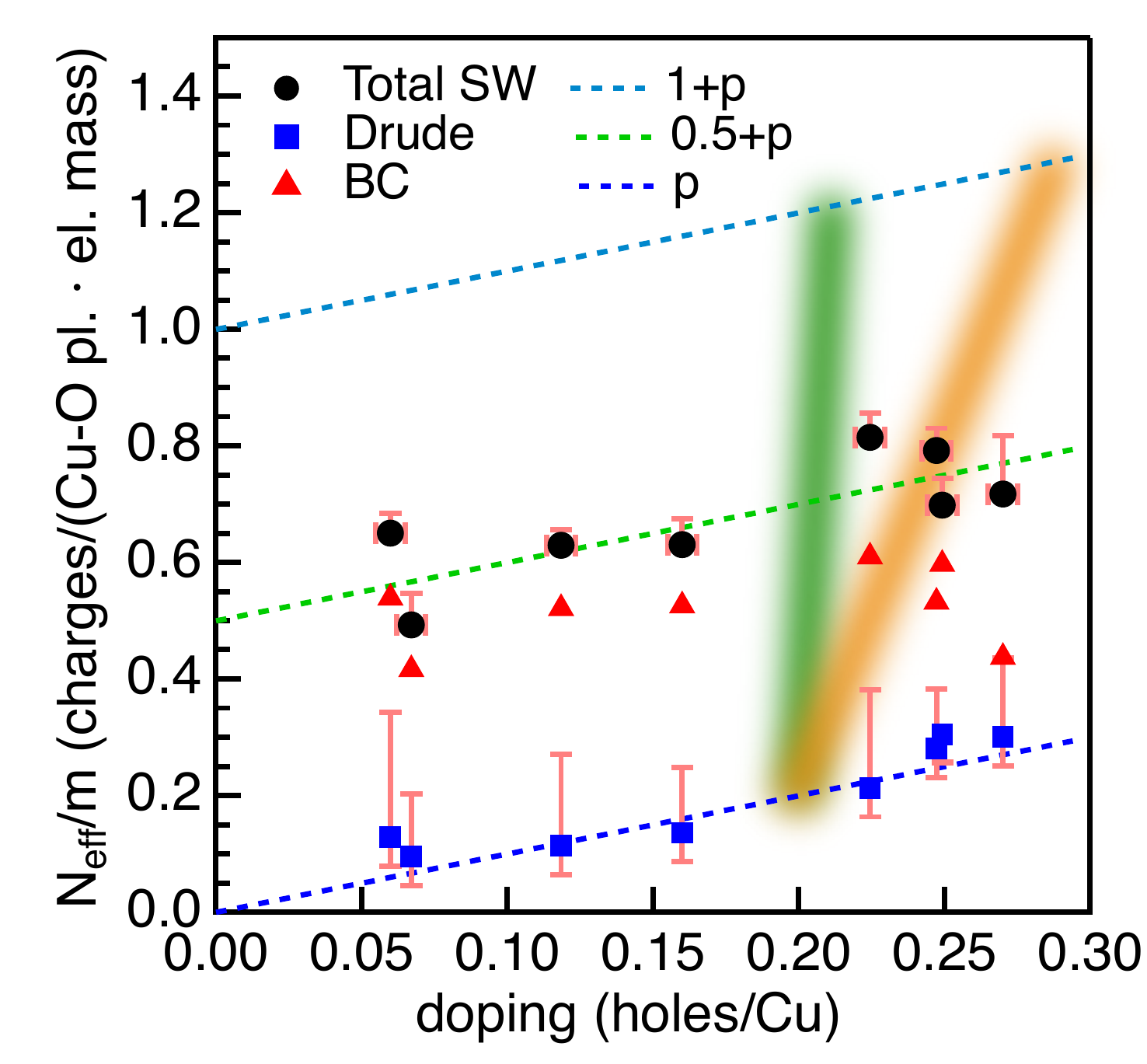}
\caption{Effective carrier density at room temperature. Solid black circles are obtained from the total integrated spectral weight, Eq. (\ref{fsumrule}). The vertical error bar on this quantity indicates the variation of the spectral weight with temperature. Blue squares indicate the spectral weight in the Drude peak, while red triangles indicate the contribution coming from the continuum. The green (Ref. \cite{Badoux:Nature2016}) and orange (Ref. \cite{Ayres:nature2021}) lines indicate observed changes in Hall carrier density.}
\label{neff}
\end{figure}
We start with the `conventional' approach, which is to integrate $\sigma_{1}(\omega)$ to $\hbar\omega_{c}$=\,1\,eV. The corresponding effective carrier density per plaquete, $N_{Dr}/m^{*}$, is shown in Fig. \ref{neff} by the black circles. The values shown are the room temperature values, while the vertical error bars indicate the change in $N_{Dr}/m^{*}$ over the entire temperature range. As was pointed out before \cite{Molegraaf:science2002, Marsiglio:PRB2008}, the normal state spectral weight follows an approximate $T^{2}$ decrease and the magnitude of this change is less than a few percent. 

In sections \ref{nofrillsDrude} and \ref{AntiMatcrossover} we have obtained a second estimate of $N_{Dr}/m^{*}$ through the estimation of the plasma frequency of the Drude component. As pointed out in section \ref{nofrillsDrude}, there is an error bar associated with the determination of the plasma frequency due to the presence of a second component. This conformal tail contribution to the low frequency $\sigma_{1}(\omega)$ was estimated in section \ref{AntiMatcrossover}. Fig. \ref{neff} shows $N_{Dr}/m^{*}$ obtained from the AM decomposition of section \ref{AntiMatcrossover} (blue squares). The upper error bar indicates the value obtained from fitting the low energy response with a single Drude component (the fits presented in Fig. \ref{drudefits}). 

The Drude weight is quite weakly temperature dependent and changes by at most 5 $\%$ up to room temperature for overdoped samples. With decreasing doping the temperature dependence increases somewhat, but never exceeds 15 $\%$. These estimates are irrespective of the method used to determine the Drude weight. This excludes directly claims to the effect that the carrier density would be strongly temperature dependent \cite{Barisic:NJP2019}. Focussing on the doping dependence, we find that $N_{Dr}/m^{*}$ exhibits a very smooth dependence on the doping $p$, independent of the method used to estimate $N_{Dr}/m^{*}$. Both the full integrated spectral weight and the Drude weight show in the slightly underdoped - strongly overdoped range a {\em simple linear increase with $p$}.

The next remarkable fact is that from Fig. \ref{neff}, it follows that the spectral weight in the incoherent conformal tail $D_{inc}$ (red triangles) is roughly doping independent {\em up to the highest doping levels}. This indicates that even the strongly overdoped cuprate metals are in this regard as strange as they are at optimal doping.

\subsection{The Mathiessen interpretation: the frequency dependence of the optical self energy.}\label{Matthiessencrossover}

The memory function $M(\omega)$, as discussed in section \ref{Gentransport}, encodes the fact that current relaxation processes are in general frequency dependent. The anti-Mathiessen interpretation of the low energy optical response of section \ref{AntiMatcrossover} puts the experimental determination of such frequency dependencies in a difficult spot: when the intraband response overlaps with generalized interband transitions, this automatically introduces spurious frequency dependence in $M(\omega)$. In addition, earlier works (Ref's \onlinecite{Dordevic:PRB2005,Hwang:PRB2007, Heumen:PRB2009} among many others) used the integrated spectral weight at 1 eV as a measure of the plasma frequency. However, in section \ref{spectralweight} we have seen that the AM derived plasma frequency is in fact quite a bit smaller. This difference has a profound impact on the extracted memory function, especially on the real part of $M(\omega)$. 

For a Drude type response we have (cf. Eq.'s (\ref{Memfunction}) and (\ref{AntiMatfit})),
\begin{eqnarray}
M(\omega)&=&\frac{\Omega_{p}}{4\pi}\frac{i}{\hat{\sigma}^{D}(\omega)}-\omega\\
&=&\left(\frac{\Omega_{p}}{\omega_{p}}-1\right)\omega+i\frac{\Omega_{p}}{\omega_{p}}\Gamma_{Dr}
\end{eqnarray}
where $\Omega_{p}$ is the plasma frequency determined for example from the integrated spectral weight. We see that errors in the determination of the spectral weight (i.e. $\Omega_{p}\neq\omega_{p}$), leads to a spurious contribution in the real part of the memory function, while the imaginary component only changes by an overall scale factor. On the other hand, when the optical response is dominated by a power law contribution to the conductivity, with $\sigma \sim (i\omega)^{-\alpha_{\sigma}}$, it follows that,
\begin{eqnarray}
M_1 (\omega) & = & \frac{\Omega_{p}^{2}}{D_{inc}}\sin (\pi \alpha/2 ) | \omega |^{\alpha} -\omega \nonumber \\ 
M_2 (\omega) & = & \frac{\Omega_{p}^{2}}{D_{inc}}\cos (\pi \alpha/2 ) | \omega |^{\alpha}
\label{conformalM}
\end{eqnarray}
We see that in contrast to the Drude result, there always is a $\omega$ term in the real part of the memory function, irrespective of the chosen $D_{inc}$ that, for $\alpha_{\sigma}\,<\,1$, dominates at large $\omega$. 
 
\begin{figure}
\includegraphics[width=0.95\columnwidth]{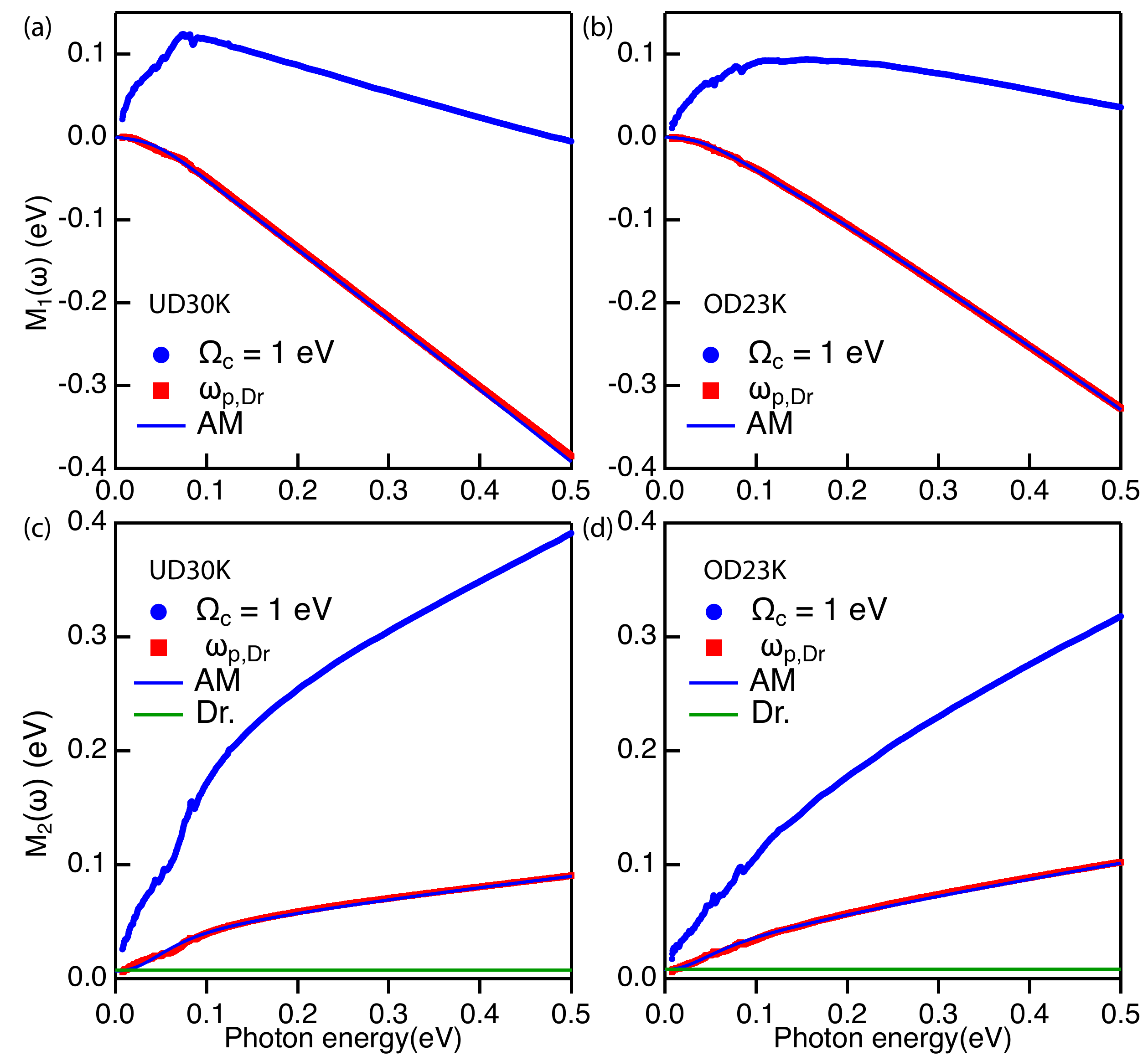}
\caption{Real (a,b) and imaginary (c,d) components of the memory function ($M(\omega)$) at 40 K. The blue circles correspond to $M(\omega)$ determined using the integrated spectral weight, while the red squares correspond to $M(\omega)$ determined using the plasma frequency obtained from the AM decomposition. The solid blue line indicates the memory function computed with the AM fits, while the green line in panels (c,d) corresponds to $\Gamma_{Dr.}$.}
\label{memfunc}
\end{figure}
With these preliminary observations, we now turn to the experimental memory function in Fig. \ref{memfunc}. Both the real and imaginary components are shown for two samples that are representative for underdoped (Fig. \ref{memfunc}a,c) and overdoped samples (Fig. \ref{memfunc}b,d). Two experimental results are shown for data measured at 40 K and for both cases we have subtracted the same contribution for high energy interband transitions (see section \ref{exp_details}). The first result (blue circles) shows $M_{SW}(\omega)$ determined from the integrated spectral weight, while the second (red squares; $M_{Dr}(\omega)$) is determined from the Drude weight. 

The difference is quite significant, especially for $M_1(\omega)$. $M_{1,SW}(\omega)$ initially increases and after a maximum goes to a negative value at high energy. $M_{1,Dr}(\omega)$ starts of almost constant and equal to zero, before approaching $M_{1}(\omega)\propto-\omega$. $M_{2,SW}(\omega)$ and $M_{2,Dr}(\omega)$ are also different, but this difference corresponds exactly to the ratio of integrated spectral weight to $\omega_{p,Dr}^{2}$ as expected. We see that when $\omega_{p,Dr}$ is used, the resulting experimental $M_{2,Dr}(\omega)$ agrees well with the Drude width (green line) determined from the fits. However, $M_{2,Dr}(\omega)$ starts to deviate from $\Gamma_{Dr}$ already around 14 meV and this can be attributed to the conformal tail contribution.

The $M(\omega)$ obtained from the AM decomposition is also shown Fig. \ref{memfunc} and accurately reflects the frequency dependence of the experimental data. We can use this to determine the origin of the remaining frequency dependence in $M_{2,Dr}(\omega)$ and find that this is mostly determined by $\Gamma_{inc}$: the larger $\Gamma_{inc}$, the larger the `step' in $M_{2,Dr}(\omega)$ around 0.1\,eV. These results make it difficult, if not impossible, to determine whether there is room for a frequency dependent momentum relaxation rate. It may be possible that there is some room for this at lower energy, but time-domain THz spectroscopy experiment seem to agree equally well with a frequency independent $\Gamma_{Dr}$\cite{Post:PRB2021}.

\section{Discussion and conclusions.}
\label{discconc}

We have attempted to push the limits with regard to extracting the maximal amount of information from the optical conductivity, resting on general principle trying to avoid any form of theoretical bias. As we emphasized in the introduction the optical conductivity (let alone the DC transport) reveals transport properties on {\em macroscopic} spatial scales that are controlled by macroscopic conservation principles. The microscopic physics enters only indirectly in determining the numbers governing the $q \rightarrow 0$ transport phenomena. A case in point is that based on this data it cannot even be decided whether the electrical transport is governed by a dilute gas of quasiparticles, as in conventional metals but with anomalous life times, or either in the highly collective hydrodynamical flow principles suggested by the `unparticle' physics of the densely entangled metals described by e.g. the AdS/CFT correspondence. 

What are the bounds that we have established? In the first place, we have much emphasized the remarkable power of the exceedingly simple {\em analytic} structure of the complex conductivity in the conformal tail regime to decompose it in two very distinct regimes. Regardless even gross differences in the basic physics, like whether it is Matthiessen- or anti-Matthiessen, just based on the way that the real- and imaginary parts of the conductivity relate to each other: a high energy branch cut regime can be distinguished from a low energy "Drude-like" response. A characteristic energy scale $\simeq 50-100$ meV is clearly discernible separating these two very different responses.

As we repeatedly emphasized, the origin of the conformal tail is plainly mysterious. All along, the intrigue with the strange metals have been spurred by physical behaviours that are unreasonably simple when viewed from a conventional theoretical perspective, the Planckian linear resistivity being point in case  \cite{Zaanen:SciPost2019,Zaanen:arxiv2021}. We perceive the `unreasonable' simplicity of the conformal tail as part of this agenda. Our analysis in Section (\ref{crossoverreg}) amplified this a bit further. The conformal tail is just referring to the fact that experimentally, both $\sigma_1(\omega)$ and $\sigma_2(\omega)$ show the same simple powerlaw behaviour where only the relative weights are different by a factor consistent with the exponent (the constant phase angle) {\em over a large frequency range}. We do know that this conformal behaviour comes to an end at the crossover scale and we showed that in the anti-Matthiessen interpretation it is quite non-trivial to reconcile this with a phase angle becoming as frequency independent as observed. 

The other news with regard to the conformal tail pertains to the doping dependence. Against our initial expectations, we were quite surprised to find out that, if anything, the conformal tail only seems to become more "perfect" when increasing doping up to a level so high that even superconductivity disappears. Strikingly, the spectral weight associated with the conformal tail appears to be roughly doping independent (Fig. \ref{neff}). Although it is straddling the effective resolution of our analysis, the only noticeable doping dependence appears to be in the exponent that appears to increase significantly as function of doping in a continuous fashion, suggesting a quantum critical phase like behaviour. 

Once again, this stresses that the `strangeness' of the metallic state in this regard apparently persists up to the highest dopings. We are of the opinion that this issue should be on the benchmark list of facts that a theory should explain -- presently, there appears to be not even a single candidate in this regard. 

Let us now turn to the low energy regime which is directly related to the DC transport properties, in the first place to the (near) linear-in-temperature resistivity. Upon lowering temperature the cuprate metals become quite good conductors. Although it is often taken for granted that it is governed by a simple Drude response our high precision analysis shows the caveats. To be sure that one is dealing which such a response one should be able to discern the low frequency plateaux followed by the $\omega^2$ fall off as in Fig.'s (\ref{drudefits})a-c. The representation in Fig.'s (\ref{drudefits}) d-h reveals precisely when this is the case: it is no longer possible to claim a response governed by a single relaxational pole at temperatures above $\simeq 150$ K. At higher temperatures the width of the Drude peak becomes of order of the crossover scale. Eventually, at the highest (room) temperature the plateaux in $\sigma_1(\omega)$ extends more or less to the crossover scale. 

As a caveat, we emphasized that on basis of the data the assignment of a discernible frequency dependence of the optical self energy (memory function) as is asserted in the generalized Drude data fitting is highly ambiguous. The frequency independent phase angle observed in our low temperature data (Section \ref{branchcut}) cannot be captured by such a perturbative form\cite{Norman:PRB2006} and the dynamical frequency range of the Drude-like peak is very limited. Stronger, as a result of the (anti) Matthiessen ambiguity even at low energy it is not possible to discriminate whether the deviations of a single relaxation time Drude response should be assigned to a broadened onset of the conformal tail or whether it is due to a frequency dependence of the optical self energy. 

This is heralding the approach to the bad metal regime where the magnitude of the (quasi) linear resistivity exceeds the maximum value that is permitted in a quasiparticle system (the Mott-Ioffe-Larkin minimal metallic conductivity). It is known at least qualitatively what happens when temperature is further increased. The roughly optical conductivity further decreases and a mid-infrared peak becomes visible (for a compilation, see Ref. \onlinecite{Delacatrez:scipost2017}). It may well be that this can be associated with the peak terminating the conformal tail: a further systematic study of the optical condcutivities at very high temperatures would be desirable. 

It appears to be unambiguous that the low temperature narrow Drude response is eventually controlled by {\em long lived total momentum}, revealing a momentum relaxation time of order of the Planckian dissipation time $\tau_{\hbar} = \hbar / (k_B T)$. This just means that when the fluid is set in motion it will move ballistically, initially with a constant velocity to come to a standstill after a time $\sim \tau_{\hbar}$. However, upon raising temperature one can no longer take this for granted relying on the optical conductivity information. As emphasized by Hartnoll \cite{Hartnoll:natphys2015}, such physics becomes untenable dealing with the magnitude of the resistivity realized at very high temperature in the bad metal regime. The momentum life time becomes of a microscopic magnitude and all motions become diffusional. Energy- and charge diffusion may take over the control in this incoherent regime -- Hartnoll's conjecture is that these may in turn be controlled by the Planckian time in the densely entangled fluid. Notice that the use of the word `incoherent' in this context is related but yet different from the zero density type quantum critical response we discussed in the introduction.   

One cannot be sure that the intermediate temperature regime, where the fingerprints of the Drude response are hard to trace and the resistivity is still relative small, can be captured by the single relaxation pole of the Drude response. It may well be that different relaxational channels are here at work. Asserting that we are dealing with the rapidly thermalizing, strongly coupled fluid this is a new territory that is presently explored using holographic means. In the presence of a strong microscopic background potential a hydrodynamical-like fluid with viscoelastic tendencies may show a response that is much richer than the plain-vanilla hydrodynamics realized in a homogeneous background\cite{Baggioli:arxiv2022}. A case in point is the hydrodynamics associated with a fluctuating pinned charge density wave with a response that reconstructs the evolution from a simple Drude peak at low temperature to the high temperature response dominated by the mid-IR resonance \onlinecite{Delacatrez:scipost2017, Delacretaz:PRB2017}. Notice the main short coming of this approach is that it does not shed any light on the issue that this resonance appears to be associated with the onset of the conformal tail. 

Regardless these ambiguities the division of the spectral response between the low energy Drude-like regime and the high energy conformal tail is beyond any doubt. This allows us to track with confidence the evolution of the spectral weights as function of doping shown in Fig. \ref{neff} -- these quantities are not particularly sensitive to the ambiguities we face dealing with the cross over since this affects only a small part of the dynamical range. The outcome is surprisingly simple: the weight in the Drude part as of relevance to the DC transport (blue squares) appears to show a simple proportionality to doping, {\em all the way to the strongly overdoped regime not revealing any irregularity at $p_c \simeq 0.19$}. As we already emphasized, the spectral weight in the conformal tail is rather doping independent, such that the f-sum rule spectral weight is uniformly increasing (black circles) like $0.5 + p$. Even at the highest dopings this does not seem to reach the weight expected from a large area Fermi surface characterized by the bandstructure effective mass (the $1+p$ line). 

This is in grave contrast with various estimates for the doping dependent carrier density based on Hall effect measurements.  One claim is that according to the Hall number the carrier density would either jump rather suddenly from $p$ to $1- p$ at the critical doping $p_c$ (indicated in Fig. \ref{neff} with a green area) interpreted as a sudden change of the $N_{Dr} \sim p$ in a doped Mott insulator to the expected carrier density of a conventional large Fermi surface Fermi-liquid \cite{Badoux:Nature2016,Lizaire:PRB2021}. This was challenged by later work, claiming instead a change of slope of the carrier density as function of doping at $p_c$ (orange area in Fig. \ref{neff}) \cite{Putzke:nphys2021,Ayres:nature2021}. 

A recent optical study, Ref. \onlinecite{Michon:PRR2021}, also found a smooth doping dependence of the full and Drude spectral weight in LSCO. These authors also find that the spectral weight does not change across $p^{*}$, but the authors argue that when changes in the mass enhancement are taken into account the optical results are in agreement with Hall experiments. The mystery is then that the mass enhancement factor should `coincidentally' and exactly cancel the change in the carrier density. We consider this unlikely, as the reported doping dependent mass enhancement factors reported in Ref's \onlinecite{Legros:nphys2019,Michon:PRR2021} are inconsistent with this interpretation. At low doping, the mass enhancement factor is of order $m*/m\approx$\,2.5. In LSCO the mass enhancement peaks around $p^{*}$\,=\,0.2 $m*/m\approx$\,12.5 and beyond that doping decreases again to $m*/m\approx$\,5 for the range of doping levels studied here. In addition, it should be emphasized that these estimates are all obtained from specific heat measurements at large magnetic fields and very low temperatures\cite{Michon:nature2019,Girod:PRB2021}. Invariably at the fields that can be reached in the laboratory one is still deep inside the flux liquid regime and given the gross changes that occur entering the superconductor from the strange metal as function of temperature (e.g., the cohering of the quasiparticles in the superconducting state) it is not at all clear whether these band mass estimates have anything to do with the high temperature metal\cite{Hsu:PNAS2021}. 

Regardless, the changes in the reported mass are large enough that we should resolve a non-monotonic doping dependence in the doping range studied here. The opposite is also true: if we combine the mass enhancement factor with the Hall density reported in Ref. \onlinecite{Lizaire:PRB2021}, the corresponding $n/m$ is much smaller than what is observed in the optical spectral weight and in addition still has a significant doping dependence. Therefore, by measuring the carrier density directly using optical means, claims of a strong change in carrier density at a critical doping are hereby proven to be flawed. 

It is actually not at all a surprise that the Hall coefficient does {\em not} directly reveal the carrier density. That one has to be careful using DC properties to infer Drude parameters (e.g., the resistivity $\sim 1/ (\omega^2_p \tau$) is even more true for magneto-transport quantities. It is a matter of principle that in order for the Hall coefficient to reveal the carrier density it should be strictly {\em temperature independent}, and this is {\em never} the case in the cuprate metals. 

This principle is rooted in symmetry: the electrical field sources {\em linear} momentum and the relaxation time associated with the resistivity is therefore associated with the {\em inhomogeneity} of space -- the breaking of translational invariance. On the other hand, the Lorentz force sources {\em angular momentum} and the associated relaxation time is associated with the {\em anisotropy} of the spatial manifold. The consequence is that a-priori the Hall relaxation time ($\tau_H$) is different from the one associated with the zero magnetic field resistivity ($\tau_J$). Assuming a simple Drude transport, it follows immediately that $R_H = (1/ n) \tau_H / \tau_J$ where $n$ is the actual carrier density that in turn should be consistent with the optically determined Drude weight. 

The present generations appear to have forgotten that in the 1990's there was much attention to the fact that $\tau_H$ shows a quite different temperature dependence than $\tau_J$. It was found that the Hall angle $\theta_H = (\omega_c \tau_H) \sim 1/T^2$ in the YBCO strange metal \cite{Harris:PRB1991}, implying $R_H \sim 1/T$. It appears that the temperature exponent of $\tau_H$ is varying depending on the various families\cite{Konstantinovic:PRB2000}, but the Hall coefficient is never temperature dependent. 

 Our optical data show that at least in the good metal regime the response is captured by a Drude transport validating the use of these simple magneto-transport wisdoms. Given the irregularities of the Hall data associated with $p_c$ which cannot be due to variations in the carrier density, these should in turn shed light on the doping dependence of the $\tau_H / \tau_J$ ratio. It would be quite worthwhile to study systematically the temperature dependences of both the Hall angle and the resistivity itself, using our optically determined Drude weight to find out what is happening with this ratio over the large doping range. 

Finally, there is more revealed by magneto-transport experiments. It was recently discovered that the magneto-resistance is highly unusual in the {\em overdoped} regime, exhibiting the `Planckian quadrature' $\rho \sim \sqrt{ (k_BT)^2 + (\mu_0 B)^2 }$ scaling behaviour\cite{Ayres:nature2021}. This is conclusive evidence for the overdoped metal to be non Fermi-liquid as well, albeit of a different kind than the underdoped strange metals. On basis of a compilation of magnetotransport data it was argued that this may reveal the existence of two parallel fluids: a more normal Fermi-liquid affair and the strange metal where the latter gradually diminishes upon overdoping\cite{Berben:arxiv2022}. This was actually part of out initial motivation to have a closer look with optics. This case rests mostly on data at low temperature in the good metal regime. When the DC transport would originate in two such parallel fluids this should then be clearly visible in the optical response in the form of two different contributions in the low energy regime -- again, the Anti-Matthiessen affair where one adds up the conductivities. However, we do observe only a single Drude peak over the whole optimally doped -- overdoped regime that is responsible for $\ge 90 \% $ of the spectral weight. Departing from the assertions in Ref. \onlinecite{Berben:arxiv2022} one would expect the sum of two very different responses that gradually exchange weight as function of doping. This can clearly be excluded on basis of our data that show a single fluid response characterized by a relaxation rate $1/\tau_J \sim T^{\beta}$ where $\beta$ appears to gradually increase from $\simeq 1$ at optimal doping towards close to 2 in the strongly overdoped regime. 

In conclusion, the theme of the strange metal as realized in the cuprates has been around since the early days of high Tc superconductivity. Nevertheless, it continues to deliver surprises. In the recent era, the overdoped regime has come into focus as being strange in its own way and not at all like the return to Fermi-liquid `normalcy' as was long believed. Our systematic, high precision study of the optical conductivity reveals in first instance that the optical response of the overdoped metal appears to be a smooth continuation of the strange metal around optimal doping. It is in essence the same affair, except for the increase of the Drude spectral weight while in other regards the scaling exponents associated with the temperature dependence of the Drude relaxation rate and the conformal tail are varying in a smooth manner. Together with the highly anomalous magneto transport properties that were very recently discovered this further strengthens the perception that we are dealing with a greatly mysterious affair that likely needs physical principle of an entirely new kind for its explanation. 

\section{Acknowledgments}
This publication is part of the project Strange Metals (with project number FOM-167) which is (partly) financed by the Dutch Research Council (NWO). 

\bibliography{paper_library}
\appendix
\section{Experimental details.}\label{appA}
In this work, we focus on single layer Bi$_{2-x}$Pb$_{x}$Sr$_{2-y}$La$_{y}$CuO$_{6+\delta}$ (BSCO), featuring a single layer in the Cu-O plane. Key advantages of BSCO over other single layer cuprates are that annealing gives access to almost the entire phase diagram and that weak van der Waals bonding in between layers allows for easy cleaving. The latter point makes it ideally suited for various spectroscopy experiments as cleaving produces surfaces that are perfectly aligned along the in-plane crystallographic directions with a mirror smooth finish. What sets our experiments apart is the focus on high resolution, temperature dependent experiments aiming to achieve a high signal-to-noise ratio and densely spaced spectra with temperature. Single crystals are grown using a travelling floating zone growth method. Subsequent annealing of as-grown crystals in vacuum or under partial oxygen atmosphere was used to change the carrier concentration. The critical temperature of the samples was subsequently determined through resistivity measurements. For all samples the same experimental sequence and parameters were used. Samples are mounted on copper cones using a silver epoxy for thermal contact. An ultra-high vacuum cryostat with temperature stabilised sample position was used in all experiments. Experimental sequences start with one to three temperature cycles between 10 K and 300 K followed by in-situ evaporation of a reference material (gold, silver or aluminium). A sample spectrum is measured at room temperature immediately before evaporation, while a reference spectrum is measured immediately after the evaporation. We verified that the room temperature spectra measured in the temperature cycle were in agreement with those measured during the evaporation of the reference. The same sequence of measurements used for the sample is repeated on the reference. This procedure guarantees reproducibility of the measured temperature dependence and eliminates small systematic movements of the sample with temperature. More details are provided in Ref. \onlinecite{TytarenkoSR:2015}. Some of the data used in this study has been published previously in Ref. \onlinecite{Heumen:NJP2009} (UD0K, UD10K, OpD35 K and OD0K).
\begin{figure}
 \includegraphics[width=0.95\columnwidth]{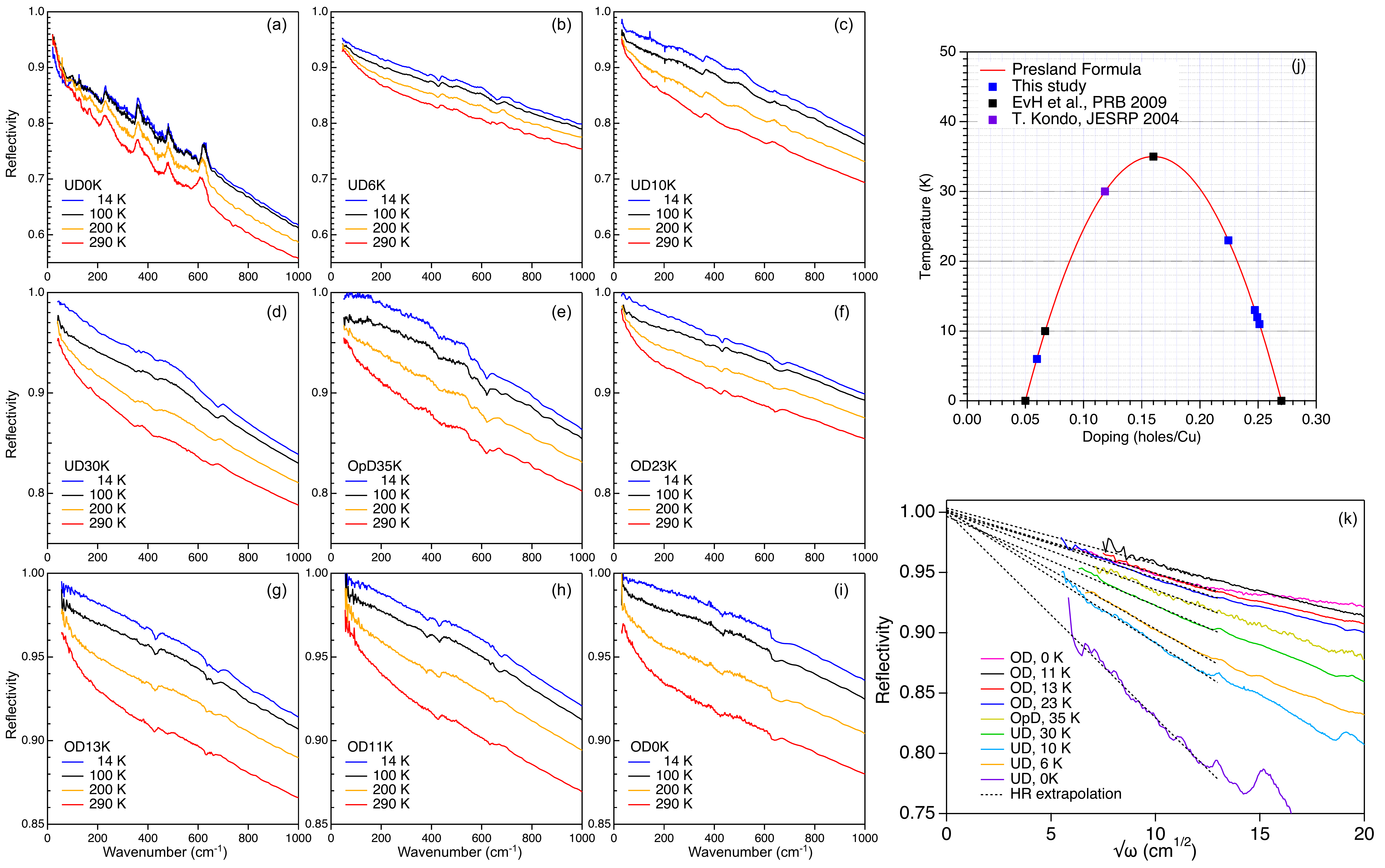}
 \caption{(Colour online) Overview of the reflectivity data used in this study. The data for the UD10K, OpD35K and OD0K crystals have been published before in Ref. \cite{Heumen:NJP2009}. Note the changes in vertical scale with increasing doping. }
 \label{fig:refl}
\end{figure}
\begin{figure*}[t][floatfix]
 \includegraphics[width=1.9\columnwidth]{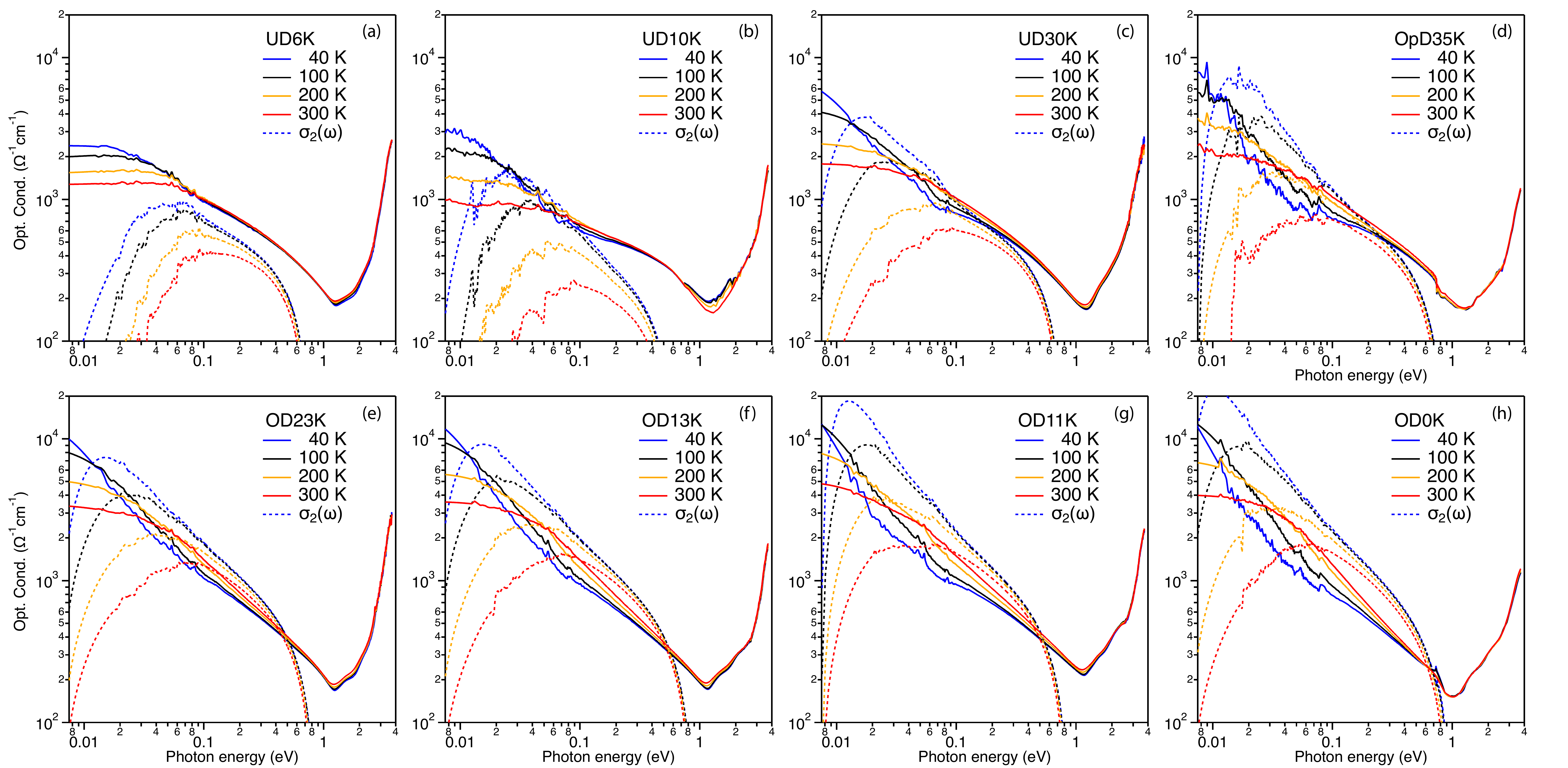}
 \caption{(a-h): Real ($\sigma_{1}(\omega)$, solid lines) and imaginary ($\sigma_{2}(\omega)$, dashed) components of the optical conductivity for a selection of temperatures and doping levels. $\sigma_{1}(\omega)$ for all doping levels and temperatures is characterized by a peak centered at zero frequency below 1 eV and interband transitions above 1 eV. }
 \label{fig:s1s2overview}
\end{figure*}
Figure \ref{fig:refl} summarises the far infrared reflectivity data, spanning the superconducting dome of the phase diagram from underdoped non-superconducting crystals to overdoped non-superconducting crystals. Starting from the overdoped side, the reflectivity is close to unity and consistent with metallic behaviour at all temperatures. With decreasing doping the reflectivity decreases and phonon modes become more prominently visible. We observe that at all doping the room temperature reflectivity appears to follow a square root law behaviour consistent with the Hagen-Rubens relation. To verify this, Fig \ref{fig:refl}k displays the reflectivity at 290 K as function of $\sqrt{\omega}$. Also shown are fits (dashed lines) with two free parameters: $R(\omega)=A-B\sqrt{\omega}$. We find that the reflectivity extrapolates to unity as expected and that the slope, $B$, increases with decreasing doping. From the fits it then follows that the slope B is set by the DC resistivity according to:
\begin{equation}
B=2\sqrt{2\varepsilon_{0}\rho_{DC}}
\end{equation}
We can thus obtain an estimate of the DC resistivity values for our crystals as function of doping. The result is shown in Fig. \ref{dcrho} and demonstrates that the values obtained from the Hagen - Rubens relation are in excellent agreement with the values obtained from transport studies \cite{Ono:PRB2003, Ayres:nature2021}. Taken together, the square root behaviour of the reflectivity and good agreement of the estimated DC resistivity, are a first indication that the low energy optical conductivity is described by a simple Drude response. 

\section{The Drude response.}\label{appB}
In this appendix we provide additional details of the measured spectra. Fig. \ref{fig:s1s2overview}(a-h) show the real and imaginary components for all samples used in this study. Starting from the UD6K sample, we see a small and broad Drude like response in $\sigma_{1}(\omega)$, which sharpens as temperature decreases. At all temperatures $\sigma_{1}(\omega)$ has a plateau at low energy followed by a fall-off above 20 meV. The mid-infrared $\sigma_{1}(\omega)$ has a clearly different second component with a hump-like feature around 0.5 eV. $\sigma_{2}(\omega)$ consists of a maximum around 0.06 - 0.1 eV, which is strongly temperature dependent. 

As doping increases, the Drude response gains spectral weight. At the same time, the Drude width narrows, as can be seen by looking at the 40 K spectra. This is also borne out by our analysis in the main paper, as presented in Fig. \ref{Drwidth}. The plateau that is visible for the UD6K and UD10K samples disappears and a narrow Drude response can be seen for higher dopings. This narrowing of the Drude response is also visible in $\sigma_{2}(\omega)$, where we see a sharp maximum emerging at low temperature. Tracking the position of this maximum with doping and temperature provides a first estimate of the changes taking place in the Drude width, demonstrating that indeed $\Gamma_{Dr}$ is the main driver for the doping dependence. 
\begin{figure*}
\includegraphics[width=1.9\columnwidth]{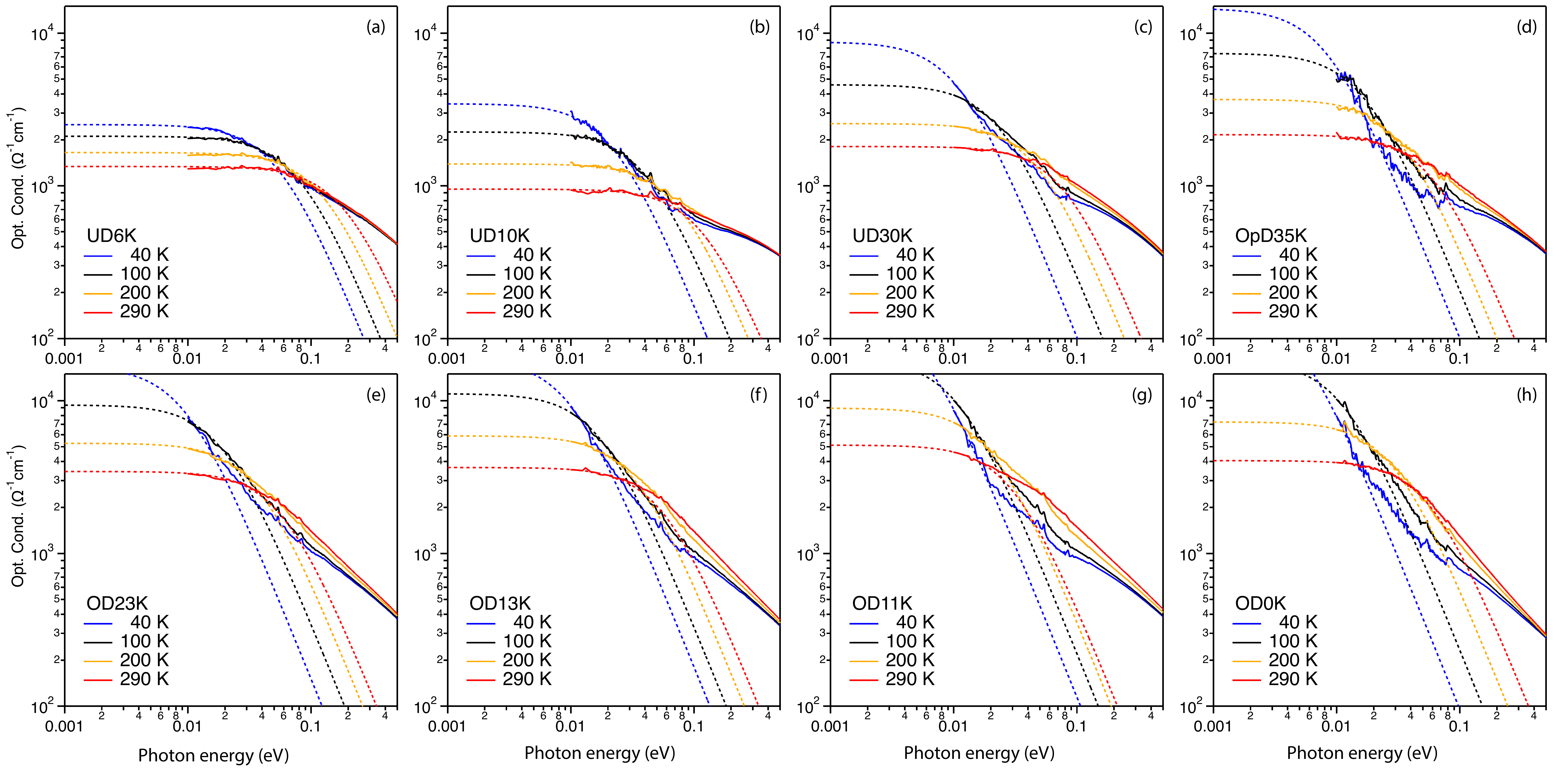}
\caption{(a-g): Real part of the optical conductivity for all measured samples, plotted on a log-log scale. Dashed lines are fits with a single Drude response of the low energy conductivity. }
 \label{sigmaoverview}
\end{figure*}
Figure \ref{sigmaoverview} presents the data underlying the analysis presented in section \ref{nofrillsDrude}. We show the fits to the low frequency optical response for all doping levels used in the analysis.

Finaly, we present the doping dependence of the scaling function, Eq. \ref{scalingfunc}. For each doping we optimize the collapse by changing the temperature coefficient $\alpha$. The resulting exponents are indicated on the vertical axis and collected in Fig. \ref{AMparams}. Note that the scaling collapse improves with increasing doping. 
\begin{figure*}
\includegraphics[width=1.95\columnwidth]{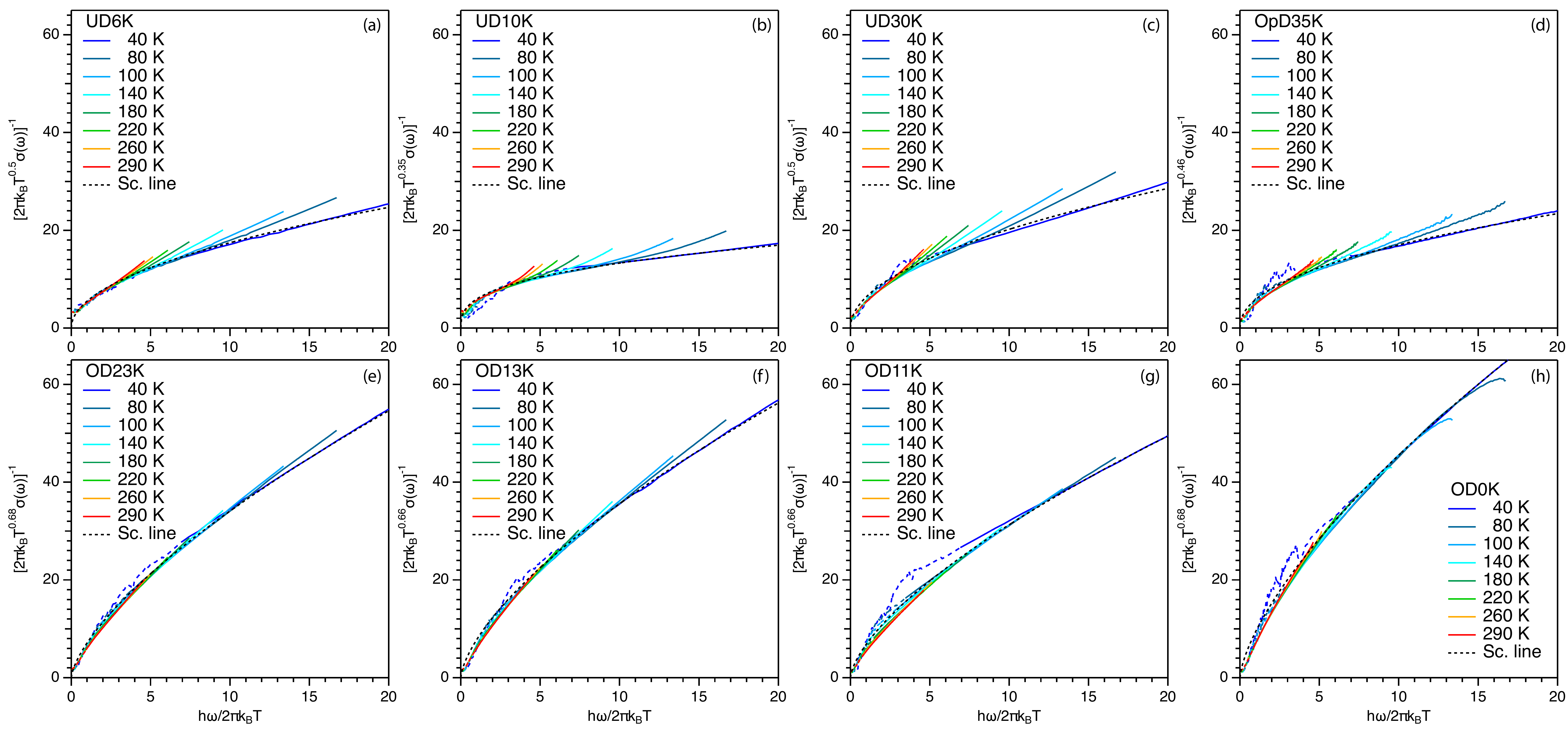}
\caption{(a-g): Temperature scaling of the real part of the optical conductivity. Dashed lines indicate the scaling relation Eq. \ref{scalingfunc}.}
 \label{phaseangleoverview}
\end{figure*}

\end{document}